\documentclass[journal,10pt]{IEEEtran}
\usepackage{graphicx}
\usepackage{caption}
\usepackage{subcaption}
\usepackage{mathtools}
\usepackage{bm,amsmath, amssymb}
\usepackage{amsmath}
\usepackage{amsfonts}
\usepackage {amssymb,amsmath}
\usepackage{cite}
\usepackage{epstopdf}
\usepackage{color}
\usepackage{microtype}
\DeclareGraphicsExtensions{.pdf,.png,.jpg}

\newcommand {\Define} {\stackrel {\Delta} {=}  }
\newcommand{\mya}{\mathrel{\overset{\makebox[0pt]{{\tiny(a)}}}{=}}}

\newtheorem{theorem}{Theorem}

\newtheorem{lemma}{Lemma}

\usepackage{cite}

\DeclareMathOperator{\sinc}{sinc}

\begin{document}
\title{\vspace{-7mm} Spectral Efficiency of OTFS Based Orthogonal Multiple Access with Rectangular Pulses}
\author{\IEEEauthorblockN{Venkatesh Khammammetti and Saif Khan Mohammed}
\IEEEauthorblockA{ \thanks{Venkatesh Khammammetti (Email: venky.crr@gmail.com) and Saif Khan Mohammed (Email: saifkmohammed@gmail.com) are with the Department of Electrical Engineering, Indian Institute of Technology Delhi (IITD), New Delhi, India. Saif Khan Mohammed is also associated with the Bharti School of Telecommunication Technology and Management (BSTTM), IIT Delhi. This work is supported by the Prof. Kishan Gupta and Pramila Gupta Chair at IIT Delhi.}}
}
\maketitle

\vspace{-22mm}
\begin{abstract}
In this paper we consider Orthogonal Time Frequency Space (OTFS) modulation based 
multiple-access (MA). We specifically consider Orthogonal MA methods (OMA) where
the user terminals (UTs) are allocated non-overlapping physical resource in the delay-Doppler (DD) and/or time-frequency (TF) domain.
To the best of our knowledge, in prior literature, the performance of OMA methods have been reported
\emph{only} for \emph{ideal} transmit and receive pulses. In \cite{OTFSMA2} and \cite{SKM_VK_MA}, OMA methods were proposed which were shown to
achieve multi-user interference (MUI) free communication with ideal pulses. Since ideal pulses are not realizable, in this paper we study the spectral efficiency (SE) performance of these OMA methods with \emph{practical rectangular pulses}.
For these OMA methods, we derive the expression for the received DD domain symbols at the base station (BS) receiver and the effective DD domain channel matrix
when rectangular pulses are used. We then derive the expression for the achievable sum SE.
These expressions are also derived for another well known OMA method where guard bands (GB) are used to reduce MUI (called as the GB based MA methods) \cite{Hadani_MA}.
Through simulations, we observe that with rectangular pulses the sum SE achieved by the method in \cite{SKM_VK_MA} is almost invariant of the Doppler shift and is
higher than that achieved by the methods in \cite{Hadani_MA, OTFSMA2} at practical values of the received signal-to-noise ratio.
\end{abstract} 

\section{Introduction}
Next generation wireless communication systems (Sixth Generation (6G) and Beyond Fifth Generation (5G)) are expected
to provide reliable communication at high data rates even for high mobility scenarios (e.g., high speed train, aircraft-to-ground
communication, etc) \cite{Paper6G, ITUIMT2020}. Current wireless communication systems are based on multi-carrier modulation
(e.g., Orthogonal Frequency Division Multiplexing (OFDM) in 3GPP 5G New Radio), whose performance degrades in high mobility
scenarios due to mobility induced Doppler spread \cite{OFDMDoppler}.

Recently, Orthogonal Time Frequency Space (OTFS) modulation has been introduced, where the information symbols are embedded
in the delay-Doppler (DD) domain  \cite{Hadani2017_Conf, Hadani2016, Hadani2018}, unlike multi-carrier systems where they are embedded in the Time-Frequency (TF) domain.
It has been shown that OTFS modulation is robust to mobility induced Doppler spread when compared to OFDM based systems \cite{Hadani2017_Conf}.
A derivation of OTFS modulation using the ZAK representation of time-domain (TD) signals has been presented in \cite{SKM2021}, where it is illustrated
that the fraction of information symbols interfered by a particular information symbol is significantly less in OTFS modulation when compared to the
fraction of sub-carriers interfered by information transmitted on a particular sub-carrier in OFDM. In other words, the effective DD domain channel
is sparser when compared to the effective TF domain channel.
Due to the sparse DD domain channel in OTFS modulation, low-complexity joint detection of information symbols can be performed in the DD domain \cite{channel, Emanuele2, lmpa, OTFSMP, mimootfs, OTFSBayes}. For DD domain detection, the estimation of the effective DD domain channel is also required \cite{ChEstOTFS1, ChEstOTFS2}.

Multi-user downlink/uplink communication based on OTFS modulation has been considered in \cite{OTFSDL1, Hadani_MA, OTFSMA1, OTFSMA2, OTFSMA3, SKM_VK_MA}.
A Non-orthogonal MA (NOMA) method has been proposed in \cite{OTFSMA1} where the same physical resource is used to communicate with
a high mobility user terminal (UT) using OTFS modulation
and also to communicate with low-mobility UTs using OFDM. However, due to non-orthogonality, the method in \cite{OTFSMA1}
can result in significant multi-user interference (MUI) for all UTs.
A low-complexity multi-user precoder has been proposed for OTFS based massive MIMO downlink systems in \cite{OTFSDL1}, where each UT
is allocated the entire physical resource and the proposed precoder is shown to result in vanishing MUI with increasing number of BS antennas.
In \cite{OTFSMA3}, a MA method has been proposed for OTFS based massive MIMO systems, where the UTs are allocated angle-domain resources
so as to ensure that there is not much MUI. This method would however be effective only for massive MIMO systems and not when the BS and UTs have few antennas.

OTFS based Orthogonal MA (OTFS-OMA) methods have been proposed in \cite{Hadani_MA, OTFSMA2, SKM_VK_MA}, where
the UTs are allocated non-overlapping resources in the DD and/or the TF domain.
In \cite{Hadani_MA}, DD domain MA methods have been proposed where guard bands are used between the resources allocated to different UTs in the DD domain,
so as to reduce MUI. However, the spectral efficiency (SE) performance of these guard band (GB) based MA methods is degraded as no information is transmitted in the guard band.
In \cite{OTFSMA2}, a non-GB based MA method is proposed where the UTs are allocated non-overlapping resources in the TF domain in an interleaved manner (subsequently referred to as ITFMA).
Ideal pulses satisfying the bi-orthogonality condition have been considered in \cite{OTFSMA2} and for which it is shown that there is no MUI.
However, such ideal pulses are not \emph{realizable} in practice. With non-ideal pulses, delay and Doppler spread would lead to \emph{severe}
MUI in the TF domain. 

Recently, in \cite{SKM_VK_MA}, a non-GB based MA method was proposed where the UTs were allocated non-overlapping resources interleaved in the DD domain in such a way that
the corresponding TF signal for each UT is restricted to a small region of the TF domain (subsequently referred to as IDDMA). The UTs are then allocated non-overlapping resources in the TF domain.
With ideal pulses, this results in MUI free communication which is shown to achieve higher SE than the
GB based MA methods. However, in \cite{SKM_VK_MA}, the impact of using \emph{practical} pulses on the SE performance of the IDDMA method
is not known.

As rectangular transmit and receive pulses have typically been used for OTFS modulation \cite{channel, PracPulseOTFS},
in this paper we analyze the sum SE performance of the IDDMA, ITFMA and the GB based MA methods, when practical rectangular pulses are used at both the
single-antenna UTs and the single-antenna BS. To the best of our knowledge, this is the first paper to comprehensively study the sum SE performance of IDDMA, ITFMA and GB based MA methods
with practical rectangular pulses. The novel contributions of this paper are as follows:
\begin{itemize}
\item In Section \ref{Section-4}, we derive the expression for the received DD domain symbols at the BS for the IDDMA, ITFMA and the
GB based MA methods when practical rectangular pulses are used. We also derive
the non-trivial expressions for the effective DD domain channel matrices for these MA methods. The expression for the achievable sum SE of these methods is also derived.
\item Through numerical simulations in Section \ref{Num_simulations}, we study the sum SE performance of
these MA methods with practical rectangular pulses. This study reveals important insights on the impact of DD/TF domain resource allocation strategy on the achieved sum SE. It is observed that for practical values of the received signal-to-noise ratio (SNR), the sum SE performance of the IDDMA method (where resource is interleaved in the DD domain and is contiguous in the TF domain) is
\emph{better} than that of the ITFMA method (where resource is interleaved in the TF domain) and the GB based methods (where guard bands are used in the DD domain).

\item In Section \ref{Num_simulations} it is observed that with rectangular pulses, the sum SE achieved by the IDDMA method is \emph{almost invariant} of the maximum
Doppler shift, whereas that achieved by the ITFMA method \emph{decreases monotonically} with increasing Doppler shift.

\item For the IDDMA method it is observed that the gap between the per-UT SE achieved with rectangular pulses and that achieved
with ideal pulses is small irrespective of the number of UTs. 
\end{itemize}

{\em Notations}: For $x \in {\mathbb R}$, $\lfloor x \rfloor$ is the greatest integer smaller than or equal to $x$.
${\mathbb E}[\cdot]$ denotes the expectation operator. ${\mathcal C}{\mathcal N}(0, \sigma^2)$ denotes
the circular symmetric complex Gaussian distribution with variance $\sigma^2$. For any
matrix ${\bf A}$, $\vert {\bf A} \vert$ denotes its determinant and ${ A}[p,q]$ denotes the
element in its $p$-th row and $q$-th column. For any two positive integers $p$ and $q$,
$( p )_q$ denotes the smallest non-negative integer which is congruent to $p$ (modulo $q$).          
 
 \section{system model}
	\label{sys_model}
	We consider an uplink MA channel where $Q$ single-antenna UTs communicate with a single-antenna BS.
	Let $s_q(t)$ denote the TD signal transmitted by the $q$-th UT ($q=0,1,\cdots, Q-1$). Also, let $h_q(\tau,\nu)$ be the delay-Doppler representation
	of the wireless channel between the BS and the $q$-th UT. The TD signal received at the BS is then given by \cite{BelloPaper}
	\begin{equation}\label{Rx_time_signal}
		r(t)=\sum_{q=0}^{Q-1} \iint h_{q}(\tau, \nu) s_{q}(t-\tau) e^{j 2 \pi \nu(t-\tau)} d \nu d \tau \, + \, w(t)
	\end{equation}
	where $w(t)$ is the Additive White Gaussian Noise (AWGN) at the BS with power spectral density $N_0$. Just as in \cite{Hadani2018} and \cite{mimootfs}, in this paper also we consider
	the delay-Doppler representation of the wireless channel between the $q$-th UT and the BS to be 
		\begin{equation}\label{h_repre}
		h_{q}(\tau, \nu)=\sum_{i=1}^{p_{q}} h_{q, i} \delta\left(\tau-\tau_{q, i}\right) \delta\left(\nu-\nu_{q, i}\right)
	\end{equation}
	where $\delta(\cdot)$ is the Dirac-delta impulse signal and $p_{q}$ is the number of propagation paths. Further, $h_{q, i}$, $\tau_{q,i}$ and $\nu_{q,i}$ are respectively the complex channel gain, the delay and the Doppler shift of the $i$-th path between the $q$-th UT and the BS. Further, we assume that
	\begin{equation}\label{max_tau_nu}
		0 \leq \tau_{q, i} < T\,,\,\left|\nu_{q, i}\right| \leq \nu_{\max }
	\end{equation}
	where $\nu_{\max}$ is the maximum possible Doppler shift of any channel path and is less than $\Delta f$.
	In this paper, we consider OTFS modulation based communication, where information symbols are embedded in the DD domain. 
	Therefore, next in Section \ref{subsecOTFSGrid}, Section \ref{subsecOTFSMod} and Section \ref{subsecOTFSDemod}, we recall the basics of OTFS modulation and demodulation.
	\subsection{OTFS Resource Grid}
	\label{subsecOTFSGrid}
	 In OTFS modulation, the information symbols of each UT are mapped on to the DD resource grid (RG). Each RG is $T$ seconds wide along the delay domain and $\Delta f=1/T$ Hz wide along the Doppler domain. The RG is sub-divided into $M$ equal parts along the delay domain and $N$ equal parts along the Doppler domain. The discretized RG in DD domain is represented as \cite{Hadani2018},
	\begin{equation}
	\Lambda_{\mbox{\tiny{DD}}} \hspace{-1mm}= \hspace{-1mm}\left\{\hspace{-1mm}\left(\frac{k}{NT},\hspace{-1mm}\frac{l}{M\Delta f}\right)\hspace{-1mm}, k=0,\dots,N-1, l=0, 
	\dots, M-1\right\}\hspace{-1mm}. \nonumber
	\end{equation}
        Each element $\left(\frac{k}{NT},\hspace{-1mm}\frac{l}{M\Delta f}\right)$ in the  discretized DD resource grid (DDRG) is uniquely identified by $(k,l)$ and is called as the $(k,l)$-th delay-Doppler Resource Element (DDRE) where $k$ is the index of this DDRE along the Doppler domain and $l$ is its index along the delay domain. 
	In OTFS modulation, DD domain information symbols in the DDRG are transformed to TF domain symbols in the TF domain RG (TFRG). The discretized RG in the TF domain is represented as
	\begin{equation}
		\Lambda_{\mbox{\tiny{TF}}} = \left\{\left(nT,m\Delta f\right), n=0,\dots,N-1,m=0,\dots, M-1\right\}. \nonumber
	\end{equation}
	Each element $\left(nT,m\Delta f\right)$ in the discretized TFRG is uniquely identified by $(n,m)$ and is called as the $(n,m)$-th Time-Frequency Resource Element (TFRE) where $n$ is the index of this TFRE along the time domain and $m$ is its index along the frequency domain. A TFRG of duration $NT$ and bandwidth $M\Delta f$ is called as an OTFS frame.
	
	\subsection{OTFS Modulation}
	\label{subsecOTFSMod}
	Let ${\Tilde x}_{q}[k,l]$ represent the information symbol of the $q$-th UT which is mapped on to the $(k,l)$-th DDRE. After mapping, DD domain information symbols are transformed to the TF domain using the Inverse Symplectic Finite Fourier Transform (ISFFT) \cite{Hadani2017_Conf} which is given by
	\begin{eqnarray}\label{OTFS_mod}
		X_{q}[n, m] & = & \frac{1}{N M} \sum_{k=0}^{N-1} \sum_{l=0}^{M-1} {\Tilde x}_{q}[k, l] e^{-j 2 \pi\left(\frac{m l}{M}-\frac{n k}{N}\right)}, \nonumber \\
		& & \hspace{-7mm} n=0,1,\dots, N-1, \, m=0,1,\dots, M-1  
	\end{eqnarray}  
	where $X_{q}[n,m]$ represents the $(n,m)$-th TF domain symbol of the $q$-th UT which is mapped to the $(n,m)$-th TFRE. Next, these TF symbols are converted to the TD transmit signal $s_{q}(t)$ by using the Heisenberg transform \cite{Hadani2018}, i.e.
	\begin{equation}\label{Tx_signal}
		s_{q}(t)=\sum_{m=0}^{M-1} \sum_{n=0}^{N-1} X_{q}[n, m] g_{t x}(t-n T) e^{j 2 \pi m \Delta f(t-n T)}
	\end{equation}
	where $g_{tx}(t)$ is the transmit pulse.
	To avoid time-domain interference between two consecutive OTFS frames, the last $\tau_{max} \Define \max_{q,i} \tau_{q,i}$ seconds
	of $s_q(t)$ (i.e., $s_q(t), t \in [NT - \tau_{max} \,,\, NT) $) is cyclically prefixed to the start of the OTFS frame (i.e., the last part is also transmitted in $t \in [-\tau_{max} \,,\, 0)$).  
	\subsection{OTFS Demodulation}
	\label{subsecOTFSDemod}	
	By substituting (\ref{Tx_signal}) and (\ref{h_repre}) in (\ref{Rx_time_signal}), the received TD signal $r(t)$ at the BS 
	is given by
	\begin{eqnarray}
		\label{Rx_TD_signal_1}
		\nonumber r(t) \hspace{-3mm} &=& \hspace{-3mm} \sum_{q=0}^{Q-1}\sum_{i=1}^{p_q}\sum_{m=0}^{M-1}\sum_{n=0}^{N-1}h_{q,i}X_{q}[n,m]g_{tx}\big(t-\tau_{q,i}-nT\big)  \nonumber \\
		& & \hspace{7mm} e^{j2\pi m\Delta f (t-\tau_{q,i})}e^{j2\pi \nu_{q,i} (t-\tau_{q,i})}  + w(t).
	\end{eqnarray} 
	By using the Wigner transform, the TD signal received at the BS, $r(t)$, is transformed back to the TF domain. The resulting received TF domain symbols are given by 
	\begin{eqnarray}\label{Rx_TF_signal}
		Y[\Tilde n, \Tilde m] & = & \int g_{r x}^{*}(t-\Tilde n T) \, r(t)  \, e^{-j 2 \pi \Tilde m \Delta f(t-\Tilde n T)}  \, dt \nonumber \\
		& & \hspace{-7mm} {\Tilde n}=0,1,\dots, N-1, \, {\Tilde m}=0,1,\dots, M-1  
	\end{eqnarray}
	where $g_{rx}(t)$ is the receive pulse. The TF symbols $Y[\Tilde n, \Tilde m]$ are transformed back to the DD domain by using the Symplectic Finite Fourier Transform (SFFT), i.e.,
	\begin{eqnarray}\label{Rx_DD_signal}
		\widehat{x}[k, l]=\sum_{\Tilde n=0}^{N-1} \sum_{\Tilde m=0}^{M-1} Y[\Tilde n, \Tilde m]  \, e^{j 2 \pi\left(\frac{\Tilde m l}{M}-\frac{\Tilde n k}{N}\right)}.
	\end{eqnarray}
	In \cite{Hadani2017_Conf}, ideal transmit and receive pulses are assumed which satisfy the bi-orthogonal condition. i.e.,
	\begin{eqnarray}
		\label{Bi-ortho}
		\int g_{rx}^{*}(t-nT) g_{tx}(t) e^{-j 2 \pi m \Delta f (t - nT)} dt  =  \delta(m) \delta(n)
	\end{eqnarray}
	Prior work done on OMA methods assumes ideal transmit and receive pulses satisfying the bi-orthogonality condition in (\ref{Bi-ortho}) \cite{OTFSMA2,SKM_VK_MA}. However, since bi-orthogonal pulses are not realizable, in the next section we derive the expressions for the sum SE of the IDDMA, ITFMA and the GB based MA methods when practical rectangular pulses are used.

	\section{OTFS Based MA with Rectangular Pulses}
    \label{Section-4}
    The transmit and receive rectangular pulses are given by 
    \vspace{-2mm}
    \begin{eqnarray}
    	\label{rect_pulses}
    	g_{tx} = g_{rx} = \begin{cases}
    		\frac{1}{\sqrt{T}}, & 0<t<T \\
    		0, & otherwise 
    	\end{cases}.
    \end{eqnarray}
    As these rectangular pulses do not satisfy the bi-orthogonal condition in (\ref{Bi-ortho}), it results in inter-carrier interference (ICI), inter symbol interference (ISI) and MUI at the receiver. All these interferences degrade the system performance. Therefore, in this section, we derive the SE performance of the IDDMA, ITFMA and the GB based MA methods when rectangular pulses
    are used.
    
   	\subsection{Interleaved DD MA (IDDMA) with rectangular pulses}
	\label{iddmasectionref}
    In this MA method, there are $Q=g_{1}g_{2}$ UTs and the information symbols of each UT are allocated DDREs which are  spaced  apart by $g_{1}$ DDREs along the delay domain and by $g_{2}$ DDREs along the Doppler domain \cite{SKM_VK_MA}. As delay and Doppler domain are sub-divided into $M$ and $N$ equal parts respectively, $g_{1}$ and $g_{2}$ divide $M$ and $N$ respectively. 	The $q$-th UT $(0\leq q \leq g_{1}g_{2}-1)$ is allocated the DDREs in the set \cite{SKM_VK_MA} 
    \begin{eqnarray}\label{Set_Rq}
    	\mathcal{R}_{q} &\Define & \Big\{(k, l) \mid k=\left\lfloor q / g_{1}\right\rfloor+g_{2} u, l=(q)_{g_{1}}+g_{1} v \nonumber \\
    	& &0 \leq u<N / g_{2}, 0 \leq v<M / g_{1}\Big\}.
    \end{eqnarray}
    \begin{figure}[h!]
    	\vspace{-2mm}
    	\centering
    	\includegraphics[width=0.70\linewidth]{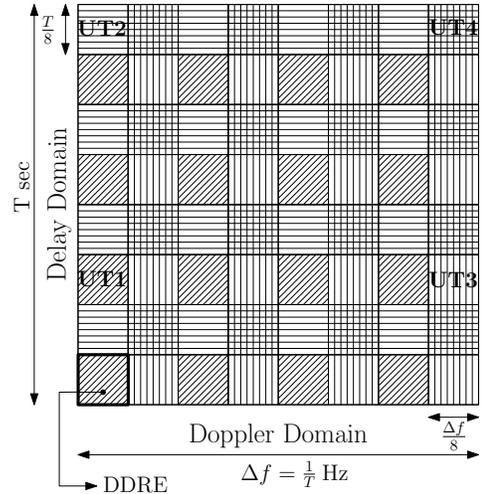}
    	\caption{Allocation of DDREs to the UTs in the IDDMA method ($M = N = 8$, $Q = 4$, $g_1 = g_2 = 2$) \cite{SKM_VK_MA}.}
    	\label{figm1}
    	\vspace{-5mm}
    \end{figure}	
    
\begin{figure}[h!]
	\vspace{-2mm}
	\centering
	\includegraphics[width=0.65\linewidth]{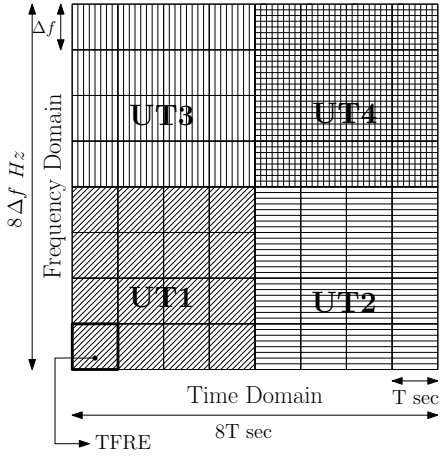}
	\caption{ Allocation of TFREs to the UTs in the IDDMA method ($M = N = 8$, $Q=4$, $g_1 = g_2 = 2$) \cite{SKM_VK_MA}.}
	\label{fig0}
	\vspace{-3mm}
\end{figure}

    In Fig. \ref{figm1}, we have shown $\mathcal{R}_{q} (q=0,1,2,3)$, for $Q=4$ UTs and $g_{1}=g_{2}=2$ (as the information symbols are interleaved in the DD domain, we refer to this method as the IDDMA method). Due to the regular spacing of DD domain information symbols, the corresponding TF symbols of each UT can be restricted to a $(NT/g_{2}) sec \times (M\Delta f/g_{1}) Hz$ region of the TFRG. As each of these TF regions is $(1/g_{1}g_{2})$-th fraction of the total TFRG and there are exactly $Q=g_{1}g_{2}$ UTs, we allocate non-overlapping continuous regions of the TFRG to the UTs (see the example in Fig. \ref{fig0} for $Q=4, g_{1}=g_{2}=2$). The non-overlapping allocation in the TF domain guarantees MUI free communication when transmit and receive pulses are ideal.
		
	After allocating DDREs from the set $\mathcal{R}_{q} (0 \leq q < Q$)  in (\ref{Set_Rq}), the TF transmitted symbols $X_{q}[n,m]$ of the $q$-th UT are given by \cite{SKM_VK_MA} 
{\vspace{-3mm}
\begin{eqnarray}
\label{newInvSFFT}
X_q[n,m] & = & \frac{1}{M N} \sum_{(k,l) \in {\mathcal R}_q }  \, {\Tilde x}_q[k,l] \, e^{-j 2 \pi {\Big (}   \frac{ml}{M}  - \frac{n k}{N} {\Big )}} \nonumber \\
&  \hspace{-5mm} \mya &  \hspace{-3mm} \lambda(m,n) \, \sum_{u=0}^{\frac{N}{g_2} - 1} \sum_{v=0}^{\frac{M}{g_1} -1}   x_q[u,v] \, e^{-j 2 \pi (\frac{m v}{M/g_1} - \frac{n u}{N/g_2})} \nonumber \\
&\hspace{-26mm}  & \hspace{-21mm} n=0,1,\cdots\hspace{-1mm},\hspace{-1mm}(N/g_2)-1) \,,\,   m=0,1,\cdots\hspace{-1mm}, \hspace{-1mm} (M/g_1)-1) 
\end{eqnarray} 
}
where $\lambda(m,n)  \Define  e^{-j 2 \pi (\frac{m (q)_{g_1}}{M}  - \frac{n \lfloor q/g_1 \rfloor}{N})}/(MN)$ and step (a) follows from the fact that ${\Tilde x}_q[k,l]$ is zero for all $(k,l) \notin {\mathcal R}_q$.  

The information symbols of the $q$-th UT $x_q[u,v] \Define {\Tilde x}_q[k = \lfloor q/g_1 \rfloor + g_2 u, l = (q)_{g_1}  + g_1 v ]$ are spaced apart by $g_2$ DDREs along the Doppler domain and by $g_1$ DDREs along the delay domain (see Fig.\ref{figm1}). Due to this,   
from (\ref{newInvSFFT}) it follows that for any integers $(\Tilde{t}_1,\Tilde{t}_2)$, $X_q[n + \Tilde{t}_1 (N/g_2) , m + \Tilde{t}_2 (M/g_1)]  =  e^{-j 2 \pi {\big (}\frac{\Tilde{t}_2 (q)_{g_1}}{g_1}  - \frac{\Tilde{t}_1 \lfloor q/g_1 \rfloor}{g_2} {\big )}} X_q[n,m]$, i.e., the TF modulated symbols are invariant to translations by integer multiples of $M/g_1$ and $N/g_2$ in the frequency  and time domain respectively. 
Therefore, it is sufficient that each UT is allocated a contiguous portion of the TFRG which is $NT/g_2$ seconds wide along the TD and is $M\Delta f/g_1$ Hz wide along the frequency domain.
Let the $q$-th UT be allocated the set of TFREs $\{ (m,n) \, | \, m=  {\big (} m' + (M/g_1) \lfloor q/g_2 \rfloor {\big  )} \,,\, n= (n' +(N/g_2) (q)_{g_2}) \,,\, m'=0,1,\cdots, M/g_1 -1 , n'=0,1,\cdots,N/g_2 -1\}$.
From (\ref{Tx_signal}) it  follows that the TD signal transmitted from the $q$-th UT is
\vspace{-2mm}
\begin{eqnarray}
\label{sqt_eqn}
s_q(t) & \hspace{-3mm} = &  \hspace{-5mm} \sum_{m=0}^{M/g_1-1}\sum_{n=0}^{N/g_2-1} X_q[n,m] g_{tx}{\big (} t - (n +(N/g_2) (q)_{g_2}) T {\big )}  \nonumber \\
& &  e^{j 2 \pi {\big (} m + (M/g_1) \lfloor q/g_2 \rfloor {\big  )} \Delta f {\big (} t - (n + (N/g_2) (q)_{g_2})T {\big )}}.
\end{eqnarray}
    	The expression for the received signal $r(t)$ at the BS is obtained by substituting (\ref{sqt_eqn}) and (\ref{h_repre}) in (\ref{Rx_time_signal}) and is given by 
    	\vspace{-2mm}
    	\begin{eqnarray}
			\label{rx_signal_r(t)}
			r(t) \hspace{-2mm} &=& \hspace{-2mm}  \sum_{q=0}^{Q-1}\sum_{i=1}^{p_q}\sum_{m=0}^{\frac{M}{g_{1}}-1}\sum_{n=0}^{\frac{N}{g_{2}}-1}h_{q,i}X_{q}[n,m]e^{j2\pi \nu_{q,i} (t-\tau_{q,i})} \nonumber \\
			& & \hspace{-12mm} e^{j2\pi \big(m+\frac{M}{g_{1}}\lfloor q / g_{2}\rfloor\big)\Delta f (t-\tau_{q,i})} g_{tx}\Big(t-\tau_{q,i}-\big(n+\frac{N}{g_{2}}(q)_{g_{2}}\big)T\Big) \nonumber \\
			& & \hspace{57mm} +w(t).
		\end{eqnarray}
		For the $q^{\prime}$-th UT, the BS computes the received TF symbols $Y_{q^{\prime}}[{\Tilde n},{\Tilde m}]$, which are given by \cite{SKM_VK_MA}
        \begin{eqnarray}
        \label{Yqprime}
        \ Y_{q^{\prime}}[{\Tilde n},{\Tilde m}] & \hspace{-3mm} =  & \hspace{-3mm} \int {\Big \{} g_{rx}^{*}{\big (} t - ({\Tilde n} + (N/g_2) (q^{\prime})_{g_2})T{\big )}  r(t) \nonumber \\
        & & \hspace{-19mm} e^{-j 2 \pi {\big (} {\Tilde m} + (M/g_1) \lfloor q^{\prime}/g_2 \rfloor  {\big )}  \Delta f  {\big (} t - ({\Tilde n} + (N/g_2) (q^{\prime})_{g_2})T{\big )}}  {\Big \}} dt.\nonumber \\
        & & \hspace{-24mm} {\Tilde n} = 0,1,\cdots, (N/g_2 -1) \,,\,  {\Tilde m} = 0,1,\cdots, (M/g_1 - 1).
\end{eqnarray}

	\begin{lemma}
		\label{lemma_1} 
		For rectangular transmit and receive pulses, the TF signal $Y_{q^{\prime}}[\Tilde n, \Tilde m]$ $(\Tilde n = 0,1,\cdots,(N/g_{2})-1, \Tilde m = 0,1,\cdots,(M/g_{1})-1)$ can be expressed in terms of the TF symbols transmitted by all the UTs, i.e.  
	    \begin{eqnarray}
			\label{Rx_TFRG_simplified_exp_rect}
		\nonumber Y_{q^{\prime}}[\Tilde n, \Tilde m]\hspace{-3mm} &=& \hspace{-3mm} 
        \underbrace{X_{q^{\prime}}[\Tilde n,\Tilde m] H_{q^{\prime},q^{\prime},1}[\Tilde m,\Tilde n,\Tilde m]}_{\text{useful term}}  \\
        & &  \hspace{-5mm}+ \underbrace{\sum_{\substack{m=0 \\ m \ne \Tilde m}} ^{M/g_{1}-1} X_{q^{\prime}}[\Tilde n,m]H_{q^{\prime},q^{\prime},1}[m,\Tilde n,\Tilde m]}_{\text{ICI term}} \nonumber \\
       	& & \hspace{-5mm}+ \underbrace{\sum_{m=0} ^{M/g_{1}-1}\sum_{\substack{q\mid  q\equiv q' (\hspace{-3mm}\mod g_{2}) \\ 0 \le q <Q \\ q \ne q'}} X_{q}[\Tilde n,m]H_{q,q^{\prime},1}[m,\Tilde n,\Tilde m]}_{\text{MUI term}} \nonumber \\
	    & & \hspace{-5mm}+ \underbrace{\sum_{m=0} ^{M/g_{1}-1}\sum_{\substack{q\mid  q\equiv q' (\hspace{-3mm}\mod g_{2}) \\ 0 \le q <Q }}\hspace{-5mm}X_{q}[\Tilde n-1,m] H_{q,q^{\prime},2}[m,\Tilde n,\Tilde m]}_{\text{ISI term}} \nonumber\\  	    
	        & & \hspace{-16mm} + \underbrace{\sum_{m=0} ^{M/g_{1}-1}\sum_{\substack{q\mid  q\equiv( q'-1) (\hspace{-3mm}\mod g_{2}) \\ 0 \le q <Q}} \hspace{-5mm} X_{q}\left[\frac{N}{g_{2}}-1,m\right]H_{q,q^{\prime},3}[m,\Tilde n,\Tilde m]}_{\text{ISI and MUI term}} \nonumber \\
	    & & \hspace{37mm}+ \underbrace{W_{q^{\prime}}[\Tilde n, \Tilde m]}_{\text{AWGN}} \nonumber \\
            & & \hspace{-17mm} {\Tilde n} = 0,1,\cdots, (N/g_2 -1) \,,\,  {\Tilde m} = 0,1,\cdots, (M/g_1 - 1),
        \end{eqnarray}where $H_{q,q^{\prime},1}[m,\Tilde n,\Tilde m],H_{q,q^{\prime},2}[m,\Tilde n,\Tilde m]$, $H_{q,q^{\prime},3}[m,\Tilde n,\Tilde m]$ and $W_{q^{\prime}}[\Tilde n, \Tilde m]$ are defined in (\ref{Define_Hq_1}), (\ref{Define_Hq_2}), (\ref{Define_Hq_3}) and (\ref{noise_rect_11}) respectively (in Appendix \ref{proof_of_lemma_1}). Also $a\equiv b  \, (\hspace{-2mm}  \mod \, g) $ iff $g \, | \, (a-b)$. 
	\end{lemma}
	\begin{IEEEproof}
		See Appendix \ref{proof_of_lemma_1}. 
	\end{IEEEproof}
	For the $q^{\prime}$-th UT, the BS transforms the TF symbols $Y_{q^{\prime}}\left[\Tilde n, \Tilde m\right]$, ($0 \leq  q^{\prime} < Q, \Tilde n = 0,1,\cdots,(N/g_{2})-1, \Tilde m =0,1,\cdots, (M/g_{1})-1$), back into  DD domain symbols ${y}_{q^{\prime}}[{\Tilde u}, {\Tilde v}]$, ($\Tilde u  = 0,1,\cdots,(N/g_{2})-1, \Tilde v =0,1,\cdots, (M/g_{1})-1$) through SFFT given in (\ref{Rx_DD_signal}),  i.e. \vspace{-2mm}
	\begin{eqnarray}
		\label{SFFT_proposed}
		y_{q^{\prime}}[{\Tilde u}, {\Tilde v}] \hspace{-2mm} & \Define & \hspace{-3mm} \sum_{\Tilde n = 0}^{N/g_2 -1} \sum_{\Tilde m = 0}^{M/g_1 -1} Y_{q^{\prime}}[\Tilde n, \Tilde m] e^{j 2 \pi {\big (} \frac{\Tilde m \Tilde v}{M/g_1}  - \frac{\Tilde n \Tilde u}{N/g_2} {\big )}}.
	\end{eqnarray} 
	The next theorem derives an expression for the received DD domain symbols, $y_{q^{\prime}}[\Tilde{u},\Tilde{v}]$ ($0 \leq q^{\prime} < Q, \Tilde u = 0,1,\cdots,(N/g_{2})-1, \Tilde v = 0,1,\cdots,(M/g_{1})-1$), in terms of the transmitted DD domain information symbols $x_{q}[u,v]$, ($0 \leq q < Q, u = 0,1,\cdots,(N/g_{2})-1, v = 0,1,\cdots,(M/g_{1})-1)$) transmitted by the $Q$ UTs.
	\begin{theorem}
		\label{Proposed_DDRG_rect_lemma_2}
		The DD domain symbols $y_{q^{\prime}}[\Tilde{u},\Tilde{v}]$ are given by \vspace{-2mm}
	\begin{eqnarray} 
			\label{Pro_DDRG_rect_expsion_1}
			y_{q'}[\Tilde{u},\Tilde{v}] &=& \underbrace{\sum_{u=0}^{N/g_{2}-1}\sum_{v=0}^{M/g_{1}-1}x_{q'}[u,v] h_{q',q',1}[\Tilde{u},\Tilde{v},u,v]}_{\text{Useful term}} \nonumber \\
			& & \hspace{-9mm}+ \hspace{-5mm} \underbrace{\sum_{\substack{q\mid  q\equiv q' (\hspace{-3mm}\mod g_{2}) \\ q \ne q'}} \hspace{-3mm}\sum_{u=0}^{N/g_{2}-1}\sum_{v=0}^{M/g_{1}-1} \hspace{-3mm}x_{q}[u,v] h_{q,q',1}[\Tilde{u},\Tilde{v},u,v]}_{\text{MUI term}} \nonumber
	\end{eqnarray}
	\begin{eqnarray}
		\label{Pro_DDRG_rect_expsion}
		& & \hspace{-6mm}+ \hspace{-7mm} \underbrace{\sum_{q\mid(q)_{g_{2}}=(q'-1)(\bmod g_{2})} \hspace{-3mm}\sum_{u=0}^{N/g_{2}-1} \sum_{v=0}^{M/g_{1}-1} \hspace{-3mm} x_{q}[u,v] h_{q,q',2}[\Tilde{u},\Tilde{v},u,v]}_{\text{MUI term}} \nonumber \\ 
		& & \hspace{34mm} + \underbrace{w_{q'}[\Tilde{u},\Tilde{v}]}_{\text{AWGN}} \nonumber \\
		& & \hspace{-12mm} {\Tilde u} = 0,1,\cdots, (N/g_2 -1) \,,\,  {\Tilde v} = 0,1,\cdots, (M/g_1 - 1),
		\end{eqnarray}
	where $h_{q,q',1}[\Tilde{u},\Tilde{v},u,v]$, $h_{q,q',2}[\Tilde{u},\Tilde{v},u,v]$ and $w_{q^{\prime}}[\Tilde{u},\Tilde{v}]$ are defined in (\ref{define_h_qq_1}), (\ref{define_h_qq_2}) and (\ref{noise_rect_DDRG}) respectively (in Appendix \ref{proof_of_lemma_2}).
	\end{theorem} 
	\begin{IEEEproof}
	See Appendix \ref{proof_of_lemma_2}.
	\end{IEEEproof}
	For the $q$-th UT, let us organize the DD domain information symbols $x_{q}[u,v], u=0,1,\cdots, N/g_2 -1, v=0,1,\cdots, M/g_1 - 1$ into the vector $\mathbf{x}_{q}$ as
	given by (\ref{vector_definition_yxw}) (see top of next page). 
		\begin{figure*}[!t]
		\vspace{-7mm}
		\begin{eqnarray}
		\label{vector_definition_yxw}
		\nonumber \mathbf{y}_{q'} \hspace{-3mm} &\Define& \hspace{-3mm} \left[\begin{array}{c} y_{q'}[0,0] , y_{q'}[1,0] ,  \hdots , y_{q'}\left[\frac{N}{g_{2}}-1,0\right] ,  y_{q'}[0,1], \hdots , y_{q'}\left[\frac{N}{g_{2}}-1,1\right], \hdots, y_{q'}\left[\Tilde{u},\Tilde{v}\right],  \hdots , y_{q'}\left[\frac{N}{g_{2}}-1,\frac{M}{g_{1}}-1\right]\end{array}\right]^{T}, \\
		\mathbf{x}_{q} \hspace{-3mm} &\Define& \hspace{-3mm} \left[\begin{array}{c} x_{q}[0,0] ,  x_{q}[1,0] , \hdots , x_{q}\left[\frac{N}{g_{2}}-1,0\right] , x_{q}[0,1] , \hdots ,  x_{q}\left[\frac{N}{g_{2}}-1,1\right] ,  \hdots ,  x_{q}\left[u,v\right] , \hdots ,  x_{q}\left[\frac{N}{g_{2}}-1,\frac{M}{g_{1}}-1\right]\end{array}\right]^{T}, \nonumber \\
		\mathbf{w}_{q'} \hspace{-3mm} &\Define& \hspace{-3mm} \left[\begin{array}{c} \hspace{-1mm}w_{q'}[0,0] , w_{q'}[1,0] ,  \hdots ,  w_{q'}\left[\frac{N}{g_{2}}-1,0\right] ,  w_{q'}[0,1] ,  \hdots , w_{q'}\left[\frac{N}{g_{2}}-1,1\right] ,  \hdots ,  w_{q'}\left[u,v\right] ,  \hdots ,  w_{q'}\left[\frac{N}{g_{2}}-1,\frac{M}{g_{1}}-1\right]\end{array}\hspace{-1mm}\right]^{T}
		\end{eqnarray}
		\hrulefill
		\vspace*{4pt}
	\end{figure*}
	Similarly, we organize the DD domain symbols received at the BS and the DD domain AWGN samples
	into the vectors $\mathbf{y}_{q'} $ and $\mathbf{w}_{q'}$ respectively (see (\ref{vector_definition_yxw})). Then from (\ref{Pro_DDRG_rect_expsion}) it follows that 
		\begin{eqnarray}{\label{vector_notation_proposed}}
		\mathbf{y}_{q'} \hspace{-3mm} &=& \hspace{-3mm} \mathbf{H}_{q',q',1}\hspace{1mm}\mathbf{x}_{q'}+  \mathbf{z}_{q'}  \nonumber \\
		\mathbf{z}_{q'} \hspace{-3mm} &\Define& \hspace{-3mm}\sum_{\substack{q\mid  q\equiv q' (\hspace{-3mm}\mod g_{2}) \\ q \ne q'}}\hspace{-5mm}\mathbf{H}_{q,q',1} \hspace{1mm}\mathbf{x}_{q}+ \hspace{-7mm} \sum_{\substack{q\mid  q\equiv (q'-1) (\hspace{-3mm}\mod g_{2}) \\ 0\le q < Q}} \hspace{-9mm}\mathbf{H}_{q,q',2} \hspace{1mm} \mathbf{x}_{q}+\mathbf{w}_{q'}.
	\end{eqnarray}

	 In (\ref{vector_notation_proposed}), $\mathbf{H}_{q,q^{\prime},i} (i=1,2)\in \mathbb{C}^{\frac{MN}{Q}\times\frac{MN}{Q}}$ and the elements of $\mathbf{H}_{q,q^{\prime},i}$ in its $\left(\Tilde u+\Tilde v(N/g_{2})+1\right)$-th row and $\left(u+ v(N/g_{2})+1\right)$-th column is $h_{q,q',i}[\Tilde{u},\Tilde{v},u,v]$, where $h_{q,q',i}[\Tilde{u},\Tilde{v},u,v]$ is defined in (\ref{define_h_qq_1}) and (\ref{define_h_qq_2}) for $i=1$ and $i=2$ respectively in Appendix \ref{proof_of_lemma_2}. Also $\Tilde u,u = 0,1,\cdots,(N/g_{2})-1$, $\Tilde v,v = 0,1,\cdots,(M/g_{1})-1$. 
	 


	Since the additive noise $w(t)$ at the BS receiver is an AWGN having power spectral density $N_0$, from (\ref{noise_rect_11}) it follows that $W_{q^{\prime}}[\Tilde n,\Tilde m]$ ($\Tilde n=0,1,\cdots,(N/g_{2})-1,\Tilde m=0,1,\cdots,(M/g_{1})-1$) are i.i.d. circular symmetric complex Gaussian distributed random variables having variance $N_0$. From the definition of $w_{q^{\prime}}[\Tilde u,\Tilde v]$ ($\Tilde u=0,1,\cdots,(N/g_{2})-1,\Tilde v=0,1,\cdots,(M/g_{1})-1)$) in (\ref{noise_rect_DDRG}), it follows that $w_{q^{\prime}}[\Tilde u,\Tilde v]$ are i.i.d. circular symmetric complex Gaussian distributed random variables having variance $\sigma^2 \Define (MNN_0)/(g_1 g_2)$, i.e., $w_{q^{\prime}}[\Tilde u,\Tilde v] \sim \mathcal{CN}(0,\sigma^{2})$. Let $x_{q}[u,v] \sim i.i.d.$  $\mathcal{CN}(0,E_{T})$ ($u=0,1,\cdots,(N/g_{2})-1,v=0,1,\cdots,(M/g_{1})-1)$). Therefore, for a given channel realization, from (\ref{vector_notation_proposed}) it follows that $\mathbf{z}_{q^{\prime}}$ is a complex Gaussian random vector whose covariance matrix is given by 
	\begin{eqnarray}{\label{covariance_proposed}}
	    \mathbf{K}_{q'} \hspace{-2mm} &\Define& \hspace{-2mm} \mathbb{E}\left[\mathbf{z}_{q^{\prime}}\mathbf{z}^{H}_{q^{\prime}}\right] \nonumber \\
		 &=& \hspace{-2mm}\sigma^{2}\mathbf{I}+E_{T} \hspace{-5mm}\sum_{\substack{q\mid  q\equiv q' (\hspace{-3mm}\mod g_{2}) \\ q \ne q'}}\hspace{-3mm}\mathbf{H}_{q,q',1} \hspace{1mm}\mathbf{H}^{H}_{q,q',1} \nonumber \\  
		& & \hspace{5mm} + E_{T}\hspace{-8mm}\sum_{\substack{q\mid  q\equiv (q'-1) (\hspace{-3mm}\mod g_{2}) \\ 0\le q < Q}}\hspace{-5mm}\mathbf{H}_{q,q',2} \mathbf{H}^{H}_{q,q',2}.
	\end{eqnarray}
	Also from the definition of $\mathbf{z}_{q^{\prime}}$ in (\ref{vector_notation_proposed}), it is clear that $\mathbf{z}_{q^{\prime}}$ and $\mathbf{x}_{q^{\prime}}$ are statistically independent. Therefore, the SE achieved by the $q^{\prime}$-th UT is given by \cite{Tse_Viswanath}
	\begin{eqnarray}{\label{SSE_proposed}}
		R_{q^{\prime}} &=& \frac{1}{MN} \log_{2} \left|\mathbf{I}+E_{T} \mathbf{H}_{q',q',1}^{H} \mathbf{K}_{q'}^{-1} \mathbf{H}_{q',q',1}\right|.
	\end{eqnarray}
	
	   	\subsection{Interleaved TF MA (ITFMA) with rectangular pulses}
		\label{itfmasectionref}
	   	In this MA method, there are $Q=g_{3}g_{4}$ UTs and each UT sends $(MN)/(g_3 g_4)$ information symbols in the DD domain. The information
		symbols of the $q$-th UT are denoted by $s_q[k',l'], k'=0,1,\cdots, (N/g_4) -1, l'=0,1,\cdots,(M/g_3) -1$. Each information symbol is quasi-periodically repeated $g_4$ times along the Doppler domain and $g_3$ times along the delay domain \cite{OTFSMA2}. Let $\Tilde x_{q}[k,l]$ denote the
		symbol transmitted by the $q$-th UT on the $(k,l)$-th DDRE. The symbol transmitted on the $(k = k^{\prime}+d_{1}(N/g_{4}), l =  l^{\prime}+c_{1}(M/g_{3}))$-th DDRE ($d_1 = 0,1,\cdots, g_4 - 1, c_1 = 0,1,\cdots, g_3 - 1$),
		 is given by \cite{OTFSMA2}
		 
		 {\vspace{-4mm}
		 \small
	   	\begin{eqnarray}
	   	\label{Interlevd_DDRE} 
	   	\Tilde x_{q}[k^{\prime}+d_{1}(N/g_{4}),l^{\prime}+c_{1}(M/g_{3})]  &  & \nonumber \\
	   	& \hspace{-54mm}  = &  \hspace{-28mm}  s_{q}[k^{\prime},l^{\prime}] e^{j2\pi\left(c_{1}\frac{(q)_{g_{3}}}{g_{3}}-d_{1}\frac{\lfloor q/g_{3} \rfloor}{g_{4}}\right)}, \nonumber \\
		& & \hspace{-37mm} d_1 = 0,1,\cdots, g_4 - 1 \,,\,  c_1 = 0,1,\cdots, g_3 - 1. 
	   	\end{eqnarray}
		\normalsize}
		Since $k'=0,1,\cdots, (N/g_4) -1, l'=0,1,\cdots,(M/g_3) -1$, from (\ref{Interlevd_DDRE}) it is clear that each UT transmits on all the DDREs (which is unlike IDDMA).  
	   	
\begin{figure}[h!]
	   		\vspace{-2mm}
	   		\centering
	   		\includegraphics[width=0.7\linewidth]{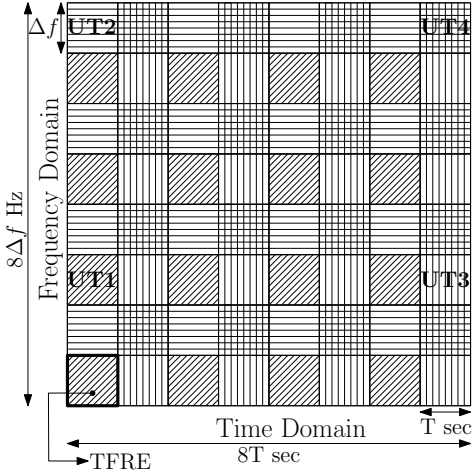}
	   		\caption{ Allocation of TFREs to the UTs in the ITFMA method ($M = N = 8$, $Q=4$, $g_3 = g_4 = 2$).}
	   		\label{fig3_int}
	   		\vspace{-5mm}
	   	\end{figure}

	   	By using the ISFFT in (\ref{OTFS_mod}), the DD domain information symbols $\Tilde x_{q}[k,l]$ (see (\ref{Interlevd_DDRE})) of the $q$-th UT are transformed to the TF domain symbols $X_{q}[n,m]$ as
	   	\begin{eqnarray}
	   	\label{Interlevd_TF}
	   	X_{q}[n,m] \hspace{-3mm} &=& \hspace{-3mm} \frac{1}{MN} \sum_{k=0}^{N-1}\sum_{l=0}^{M-1}\Tilde x_{q}[k,l]e^{-j2\pi\left(\frac{ml}{M}-\frac{nk}{N}\right)} \nonumber \\
	   	\hspace{-7mm} & \mya & \frac{1}{MN}\sum_{k'=0}^{\frac{N}{g_{4}}-1}\sum_{l'=0}^{\frac{M}{g_{3}}-1}s_{q}[k',l']e^{-j2\pi \left(\frac{ml'}{M}-\frac{nk'}{N}\right)} \nonumber \\
	   	 & & \sum_{c_1=0}^{g_3-1} e^{j 2 \pi \frac{c_1}{g_3}\left((q)_{g_3}-m\right)} \sum_{d_1=0}^{g_4-1} e^{j 2 \pi \frac{d_1}{g_4}(\lfloor q / g_3\rfloor-n)} \nonumber \\
	   	 & & \hspace{-14mm}n=0,1,2, \cdots, N-1, \,\, m=0,1,2, \cdots,M-1
	   	\end{eqnarray} where in step (a) we have used (\ref{Interlevd_DDRE}). 
	   	From (\ref{Interlevd_TF}) it is clear that $X_{q}[n,m]$ is non-zero if and only if (iff) $\left(m-(q)_{g_{3}}\right)_{g_{3}} =0$ and $(n-\lfloor q/g_3 \rfloor )_{g_{4}} = 0$ and is zero otherwise. As a result the $q$-th UT occupies only $MN/g_{3}g_{4}$ TFREs where these occupied TFREs are spaced apart by $g_{3}$ TFREs along the frequency domain and by $g_{4}$ TFREs along the time domain. As the TF domain symbols of each UT are interleaved in the TFRG, this method is termed as ITFMA (see Fig. \ref{fig3_int}). The transmitted TD signal, $s_{q}(t)$, of the $q$-th UT is then given by (\ref{Tx_signal}).
   	\begin{lemma}
   		\label{lemma3_Interlvd_TF} 
   		For the  ITFMA method with rectangular transmit and receive pulses, the received TF domain symbols $Y[\Tilde n,\Tilde m]$ at the BS are given by
   		\begin{eqnarray}
   		\label{TF_interlvd_l3}
   		Y[ \Tilde{n}, \Tilde{m}] \hspace{-3mm}&=&\hspace{-3mm}\sum_{q=0}^{Q-1}\sum_{m=0}^{M-1} \Big\{X_{q}[\Tilde{n},m] \ddot{H}_{q,1}[m,\Tilde{n},\Tilde m] \nonumber \\
   		& &   \hspace{-3mm} + X_{q}[\Tilde{n}-1,m] \ddot{H}_{q,2}[m,\Tilde{n},\Tilde m]\Big\} +W[\Tilde n,\Tilde m] \nonumber \\
		   	 & & \hspace{-14mm}\Tilde n=0,1,2, \cdots, N-1, \,\, \Tilde m=0,1,2, \cdots,M-1.
   		\end{eqnarray}
   		where $\ddot{H}_{q,1}[m,\Tilde{n},\Tilde m]$, $\ddot{H}_{q,2}[m,\Tilde{n},\Tilde m]$ and $W[\Tilde n,\Tilde m]$ are defined in (\ref{H_q_1_doppler}), (\ref{H_q_2_doppler}) and (\ref{noise_GBMA}) respectively (in Appendix \ref{proof_of_lemma3_Doplr_1}).
   	\end{lemma}
   	\begin{IEEEproof}
   		See Appendix \ref{proof_of_lemma3_Doplr_1}.
   	\end{IEEEproof}
	Due to the interleaved multiple-access in the TF domain, at the BS the TF symbols of the $q^{\prime}$-th UT are
	taken from the symbols received on the TFREs allocated to the $q^{\prime}$-th UT. Therefore,
	the received TF domain symbols for the $q^{\prime}$-th UT are given by
	    
	{\vspace{-4mm}
	\small    
	     \begin{eqnarray}
              \label{ITF_qprime}
              Y_{q^{\prime}}[\Tilde{n}, \Tilde{m}]  &  \hspace{-3mm} = \hspace{-1mm}\begin{cases}
              Y[\Tilde{n}, \Tilde{m}]  & \hspace{-3mm} , \left(\Tilde m-(q')_{g_3}\right)_{g_3}=0, (\Tilde n-\lfloor \frac{q'}{g_3} \rfloor )_{g_4} = 0 \\
              0 & \hspace{-3mm} , \mbox{\small{otherwise}}
              \end{cases}.
              \end{eqnarray}
              \normalsize}
    For the $q^{\prime}$-th UT, the BS transforms the TF symbols $Y_{q^{\prime}}[\Tilde{n}, \Tilde{m}]$ into  DD domain symbols ${y}_{q^{\prime}}[{\Tilde k}, {\Tilde l}]$, ($\Tilde k  = 0,1,\cdots,(N/g_{4})-1, \Tilde l =0,1,\cdots, (M/g_{3})-1$) through SFFT given in (\ref{Rx_DD_signal}),  i.e. \vspace{-2mm}
    \begin{eqnarray}
    \label{SFFT_Inter}
    y_{q^{\prime}}[{\Tilde k}, {\Tilde l}] \hspace{-2mm} & = & \hspace{-3mm}  \sum\limits_{\Tilde{n}=0}^{N-1} \sum\limits_{\Tilde{m}=0}^{M-1} Y_{q^{\prime}}[\Tilde{n}, \Tilde{m}] e^{j 2 \pi \left( \frac{\Tilde{m} {\Tilde l}}{M}  - \frac{\Tilde{n} {\Tilde k}}{N}  \right)} \nonumber \\
    \hspace{-2mm} &  \mya & \hspace{-3mm}   \sum_{ n^{\prime} = 0}^{N/g_4 -1} \sum_{ m^{\prime} = 0}^{M/g_3 -1} Y_{q^{\prime}}[\lfloor q^{\prime}/g_3 \rfloor + n^{\prime}g_4, (q^{\prime})_{g_3}  + m^{\prime}g_3] \nonumber \\
    & & \hspace{5mm} e^{j 2 \pi {\big (} \frac{((q^{\prime})_{g_3}  + m^{\prime}g_3)\Tilde l}{M}  - \frac{(\lfloor q^{\prime}/g_3 \rfloor + n^{\prime}g_4) \Tilde k}{N} {\big )}} \nonumber \\
    & & \hspace{-16mm}\Tilde k=0,1,2, \cdots, N/g_{4}-1, \,\, \Tilde l=0,1,2, \cdots,M/g_{3}-1,
    \end{eqnarray} 
    where step (a) follows from (\ref{ITF_qprime}).
    \begin{theorem}
    	\label{lema_interlevd} 
    	For the $q^{\prime}$-th UT ($0 \leq q^{\prime} < Q$), the received DD domain symbol $y_{q^{\prime}}[\Tilde{k},\Tilde{l}]$ ($\Tilde k = 0,1,\cdots,(N/g_{4})-1, \Tilde l = 0,1,\cdots,(M/g_{3})-1$) depends on the transmitted DD domain information symbols $s_{q}[k^{\prime},l^{\prime}]$, ($0 \leq q < Q, k^{\prime} = 0,1,\cdots,(N/g_{4})-1, l^{\prime} = 0,1,\cdots,(M/g_{3})-1)$) of all $Q$ UTs,
	and is given by
    	\begin{eqnarray}
    	\label{lema_int_exp}
    	\nonumber  y_{q'}[\Tilde{k},\Tilde{l}] &=&   \underbrace{s_{q^{\prime}}[k',l']  \, \ddot{h}_{q',q'}[\Tilde{k},\Tilde{l},k',l']}_{\text{Useful term}} \\
    	& & \hspace{-21mm} + \underbrace{\sum_{\substack{q=0 \\ q \ne q^{\prime}}}^{Q-1}\sum_{k'=0}^{\frac{N}{g_{4}}-1}\sum_{l'=0}^{\frac{M}{g_{3}}-1} s_{q}[k',l']  \, \ddot{h}_{q,q'}[\Tilde{k},\Tilde{l},k',l']}_{\text{MUI term}} + \underbrace{w_{q'}[\Tilde{k},\Tilde{l}]}_{\text{AWGN}},
    	\end{eqnarray}
    		where $\ddot{h}_{q,q'}[\Tilde{k},\Tilde{l},k',l']$ and $w_{q'}[\Tilde{k},\Tilde{l}]$ are defined in (\ref{h_mat_interlevd}) and (\ref{noise_int}) respectively in Appendix \ref{proof_of_lemma_interlevd}.
    \end{theorem}
	\begin{IEEEproof}
		See Appendix \ref{proof_of_lemma_interlevd}.
	\end{IEEEproof}
	Let the DD signal $y_{q^{\prime}}[\Tilde k,\Tilde l]$, ($\Tilde k = 0,1,\cdots, \frac{N}{g_4}-1,\,\, \Tilde l = 0,1,\cdots, \frac{M}{g_3}-1$), be arranged into the vector $\mathbf{y}_{q^{\prime}} \in \mathbb{C}^{\frac{MN}{g_{3}g_{4}}\times 1}$ such that the  $(\Tilde k+\Tilde l(N/g_{4})+1)$-th element of $\mathbf{y}_{q^{\prime}}$ is $y_{q^{\prime}}[\Tilde k,\Tilde l]$.
	Similarly let $\mathbf{w}_{q^{\prime}} \in  \mathbb{C}^{\frac{MN}{g_{3}g_{4}}\times 1}$ denote the vector of AWGN samples $w_{q^{\prime}}[\Tilde k,\Tilde l]$.
	Also let $\mathbf{{s}}_{q} \in \mathbb{C}^{\frac{MN}{g_{3}g_{4}}\times 1}$ denote the vector of DD domain information symbols transmitted by the $q$-th UT, where the $\left(k^{\prime}+ l^{\prime}(N/g_{4})+1\right)$-th element of this vector is ${s}_{q}[ k^{\prime}, l^{\prime}]$ ($k^{\prime} = 0,1,\cdots, \frac{N}{g_4}-1,\,\, l^{\prime} = 0,1,\cdots, \frac{M}{g_3}-1$).
	From (\ref{lema_int_exp}) it then follows that
	\vspace{-2mm}
	\begin{eqnarray}{\label{vector_notation_intrlvd}}
	\mathbf{y}_{q'} \hspace{-3mm} &=& \mathbf{\ddot{H}}_{q',q'} \, \mathbf{{s}}_{q'}+ \mathbf{z}_{q^{\prime}}^{\mbox{\tiny{Int}}}, \nonumber \\
	\mathbf{z}_{q^{\prime}}^{\mbox{\tiny{Int}}} &\Define& \sum_{\substack{q=0 \\ q \ne q'}}^{Q-1}\mathbf{\ddot{H}}_{q,q'} \, \mathbf{{s}}_{q}+\mathbf{w}_{q'}
	\end{eqnarray}
	where $\mathbf{\ddot{H}}_{q,q^{\prime}} \in \mathbb{C}^{\frac{MN}{g_{3}g_{4}}\times \frac{MN}{g_{3}g_{4}}}$ has $\ddot{h}_{q,q^{\prime}}[\Tilde k,\Tilde l,k^{\prime},l^{\prime}]$ as its element in its $(\Tilde k+\Tilde l(N/g_{4})+1)$-th row and $\left(k^{\prime}+ l^{\prime}(N/g_{4})+1\right)$-th column. Since $w(t)$ is AWGN with power spectral density $N_0$, from (\ref{noise_int}) it follows that $w_{q'}[k^{\prime},l^{\prime}] \sim i.i.d. \hspace{2mm} \mathcal{CN}(0,MNN_{0}/g_{3}g_{4})$. Also, let ${s}_{q^{\prime}}[k^{\prime},l^{\prime}] \sim i.i.d. \hspace{2mm} \mathcal{CN}(0,E_{T})$. From (\ref{vector_notation_intrlvd}) it then follows that the SE achieved by the $q^{\prime}$-th UT is given by \cite{Tse_Viswanath}
	\begin{eqnarray}{\label{SSE_intrlvd_final}}
	R_{q'}^{\mbox{\tiny{Int}}} &=& \frac{1}{MN} \log_{2} \left|\mathbf{I}+E_{T} \mathbf{\ddot{H}}_{q',q'}^{H} \mathbf{K}_{{q'}}^{\mbox{\tiny{Int}}^{-1}} \mathbf{\ddot{H}}_{q',q'}\right|
	\end{eqnarray}
	where $\mathbf{K}_{q^{\prime}}^{\mbox{\tiny{Int}}}  \hspace{-1mm} \Define \hspace{-1mm} \mathbb{E}\left[\mathbf{z}_{q^{\prime}}^{\mbox{\tiny{Int}}}\mathbf{z}_{q^{\prime}}^{\mbox{\tiny{Int}}^{H}}\right] = E_T  \hspace{-2mm} \sum\limits_{q=0, q \ne q'}^{Q-1}  \hspace{-2mm} \mathbf{\ddot{H}}_{q,q'} \mathbf{\ddot{H}}_{q,q'}^H   +  (MNN_0/g_{3}g_{4}) \mathbf{I}$ is the covariance matrix of $\mathbf{z}_{q^{\prime}}^{\mbox{\tiny{Int}}}$. 
\subsection{GB based MA Methods with rectangular pulses}
\label{gbmasectionref}
\begin{figure}[ht]
			\centering
			\includegraphics[width=0.65\linewidth]{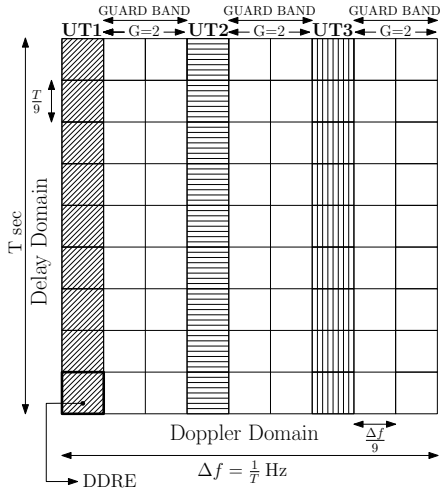}  
			\caption{Guard Band in Doppler domain based MA ($M = N = 9$, $Q = 3, G = 2$) \cite{SKM_VK_MA}.}
			\label{fig:sub-first}
			\vspace{-4mm}
	\end{figure}

\begin{figure}[ht]
	\centering
	\includegraphics[width=0.75\linewidth]{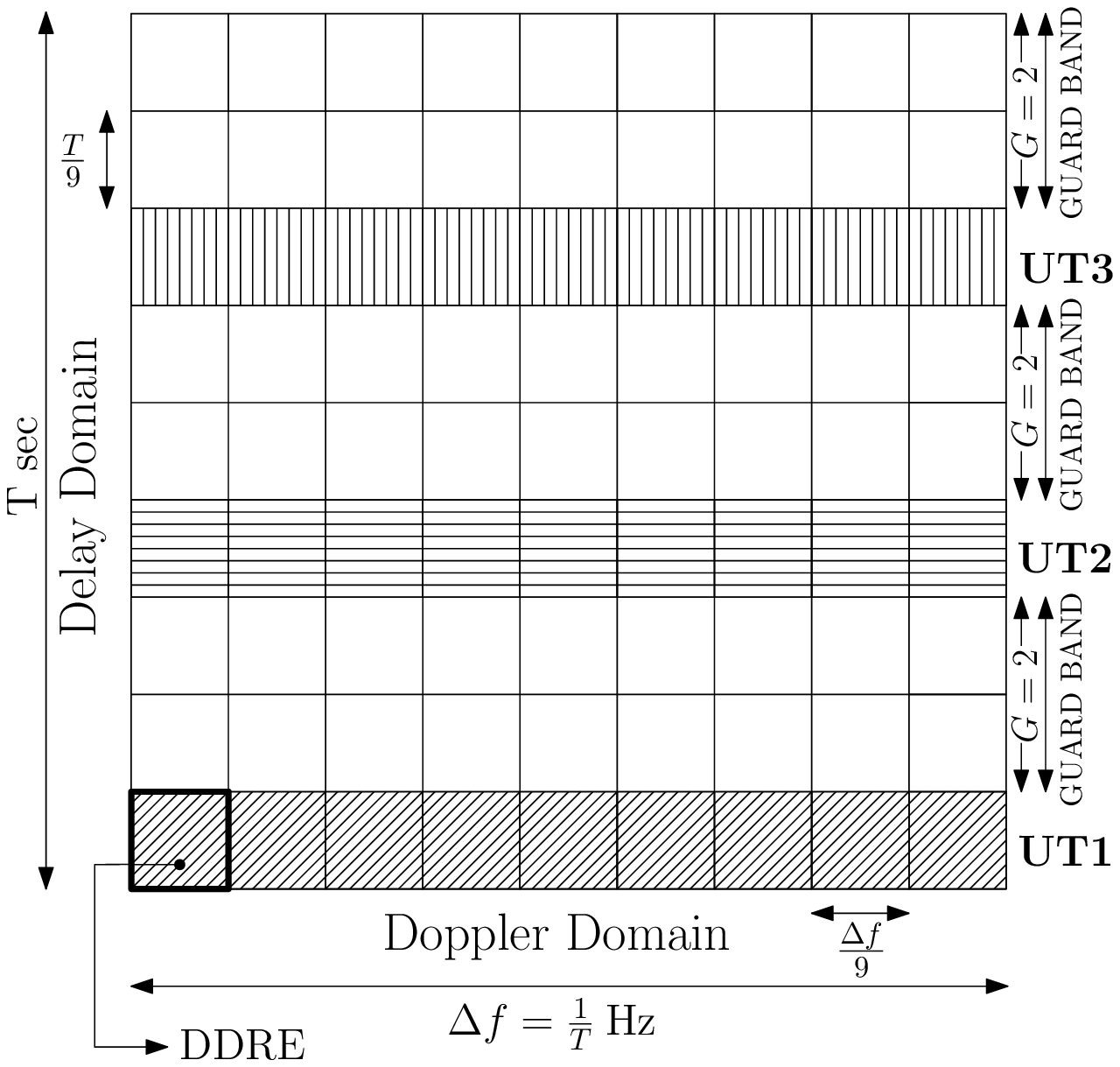}  
	\caption{Guard Band in Delay domain based MA ($M = N = 9$, $Q = 3, G = 2$) \cite{SKM_VK_MA}.}
	\label{fig:sub-second}
\end{figure}

In GB based methods, guard bands are used in the DD domain to separate the information symbols transmitted by each UT \cite{Hadani_MA}.  In the following, we specifically consider two different GB based MA methods, referred to as ``\textit{Doppler domain GB based MA}" and ``\textit{Delay domain GB based MA}". In Doppler domain GB based MA, the DDREs allocated to different UTs are separated by guard bands along the Doppler domain. With $Q$ UTs, the DDREs allocated to the $q$-th $(0 \leq q <Q)$ UT is given by the set $\mathcal{R}_{q}^{Dplr}$ as follows
\vspace{-4mm}
\begin{eqnarray}\label{Set_R_Doplr,q} 
		\mathcal{R}_{q}^{Dplr} \hspace{-2mm} &\Define & \hspace{-2mm}\Big\{(k, l) \mid q\frac{N}{Q} \leq k \leq \left(\frac{N}{Q}-G-1\right)+q  \frac{N}{Q}, \nonumber \\ 
		& &  \hspace{30mm} 0 \leq l \leq M-1 \Big\}
\end{eqnarray}
where $G$ is the guard band size in number of guard DDREs along the Doppler domain.
As an example, for $Q=3,G=2,M=N=9$, the allocation of DDREs to each UT is shown in Fig. \ref{fig:sub-first} (the unshaded areas are GBs which are not used for transmitting information).
	 \\
In Delay domain GB based MA, the DDREs allocated to different UTs are separated by guard bands along the delay domain. With $Q$ UTs, the DDREs allocated to the $q$-th $(0 \leq q <Q)$ UT is given by the set $\mathcal{R}_{q}^{Dly}$ as follows
\vspace{-3mm}
\begin{eqnarray}\label{Set_R_Dly,q}
		\mathcal{R}_{q}^{Dly} &\Define & \Bigg\{(k, l) \mid 0 \leq k \leq N-1, q \frac{M}{Q} \leq l \leq \nonumber \\
    	&& \hspace{16mm}\left(\frac{M}{Q}-G-1\right)+q  \frac{M}{Q} \Bigg\}
\end{eqnarray}
where $G$ is the guard band size in number of guard DDREs along the delay domain.
As an example, for $Q=3,G=2,M=N=9$, the allocation of DDREs to each UT is shown in Fig. \ref{fig:sub-second} (the unshaded areas are GBs which are not used for transmitting information). 

\subsubsection{Analysis of the sum SE achieved by Doppler domain GB based MA method with rectangular transmit and receive pulses} 
From (\ref{Set_R_Doplr,q}), it follows that the TF domain symbols at the $q$-th UT are given by \vspace{-2mm}
\begin{eqnarray}
\label{newInvSFFT_Doppler}
X_q[n,m] & = & \frac{1}{M N} \sum_{(k,l) \in {\mathcal R}_{q}^{Dplr} }  \, {\Tilde x}_q[k,l] \, e^{-j 2 \pi {\Big (}   \frac{ml}{M}  - \frac{n k}{N} {\Big )}} \nonumber \\
&  \hspace{-5mm} \mya &  \hspace{-3mm} \frac{1}{MN} \sum_{l=0}^{M-1}\sum_{k=q\cdot\frac{N}{Q}}^{\left(\frac{N}{Q}-G-1\right)+q\cdot\frac{N}{Q}} \hspace{-3mm} {\Tilde x}_{q}[k,l]e^{-j2\pi \left(\frac{ml}{M}-\frac{nk}{N}\right)} \nonumber \\
&\hspace{-26mm}  & \hspace{-16mm} n=0, \cdots, N-1, m=0,\cdots, M-1,
\end{eqnarray} 
where step $(a)$ follows from the fact that ${\Tilde x}_{q}[k,l]=0$ for $(k,l) \notin \mathcal{R}_{q}^{Dplr}$. The transmitted TD signal, $s_{q}(t)$, of the $q$-th UT is then given by (\ref{Tx_signal}).

For the Doppler domain GB based MA method with rectangular transmit and receive pulses, the received TF domain symbols $Y[\Tilde n,\Tilde m]$ at the BS are given by Lemma \ref{lemma3_Interlvd_TF} (with $X_q[n,m]$ given by (\ref{newInvSFFT_Doppler})).
The DD domain symbols received in $\mathcal{R}_{q^{\prime}}^{Dplr}$ are used to demodulate the information symbols transmitted by the $q^{\prime}$-th UT. The following theorem gives the expression of the DD domain symbols received in $\mathcal{R}_{q^{\prime}}^{Dplr}$ in terms of the DD domain information symbols transmitted by all UTs.
\begin{theorem}
    \label{Lemma4_Doppler_DD}
    For the $q^{\prime}$-th UT ($0 \leq q^{\prime} < Q$), the DD domain symbols $y_{q^{\prime}}[\Tilde{k},\Tilde{l}]$ received at the BS for $(\Tilde{k},\Tilde{l}) \in \mathcal{R}_{q^{\prime}}^{Dplr}$ are given by \vspace{-3mm}
    \begin{eqnarray}
			\label{Doplr_DDRG_rect_expsion}
			\nonumber  y_{q^{\prime}}[\Tilde{k},\Tilde{l}] &=&  \underbrace{\sum_{l=0}^{M-1}\sum_{k=q'\cdot \frac{N}{Q}}^{\left(\frac{N}{Q}-G-1\right)+q'\cdot \frac{N}{Q}}{\Tilde x}_{q'}[k,l] \Tilde{h}_{q',q'}[\Tilde{k},\Tilde{l},k,l]}_{\text{Useful term}} \\
			 & & \hspace{-16mm}+ \underbrace{\sum_{\substack{q=0 \\ q \ne q'}}^{Q-1}\sum_{l=0}^{M-1}\sum_{k=q\cdot \frac{N}{Q}}^{\left(\frac{N}{Q}-G-1\right)+q\cdot \frac{N}{Q}} \hspace{-7mm} {\Tilde x}_{q}[k,l] \Tilde{h}_{q,q'}[\Tilde{k},\Tilde{l},k,l]}_{\text{MUI term}} + \underbrace{w[\Tilde{k},\Tilde{l}]}_{\text{AWGN}} 
		\end{eqnarray}
		where $\Tilde{h}_{q,q'}[\Tilde{k},\Tilde{l},k,l]$ and $w[\Tilde{k},\Tilde{l}]$ are defined in (\ref{define_h_q_doplr}) and (\ref{w_k_l_Doplr}) respectively (in Appendix \ref{proof_of_Doplr_DDRG_lemma}) and $(\Tilde k,\Tilde l) \in  \mathcal{R}_{q^{\prime}}^{Dplr}$.
\end{theorem}
\begin{IEEEproof}
		See Appendix \ref{proof_of_Doplr_DDRG_lemma}.
\end{IEEEproof}
Let the DD signal $y_{q^{\prime}}[\Tilde k,\Tilde l]$, ($(\Tilde k,\Tilde l) \in \mathcal{R}_{q^{\prime}}^{Dplr}, 0 \leq q^{\prime} < Q$), be arranged into the vector $\mathbf{y}_{q^{\prime}} \in \mathbb{C}^{M\left(\frac{N}{Q}-G\right)\times1}$ such that the  ($ (\Tilde k-q^{\prime}(N/Q))+\Tilde l ((N/Q)-G)+1$)-th element of $\mathbf{y}_{q^{\prime}}$ is $y_{q^{\prime}}[\Tilde k,\Tilde l]$.
Similarly let $\mathbf{w}_{q^{\prime}} \in  \mathbb{C}^{M\left(\frac{N}{Q}-G\right)\times1}$ denotes the vector of AWGN samples $w[\Tilde k,\Tilde l]$, ($(\Tilde k,\Tilde l) \in \mathcal{R}_{q^{\prime}}^{Dplr}, 0 \leq q^{\prime} < Q$).
Also let $\mathbf{{\Tilde x}}_{q} \in \mathbb{C}^{M\left(\frac{N}{Q}-G\right)\times1}$ denote the vector of DD domain information symbols transmitted by the $q$-th UT, where the ($ ( k-q(N/Q))+ l ((N/Q)-G)+1$)-th element of this vector is ${\Tilde x}_{q}[ k, l]$, where $( k, l) \in \mathcal{R}_{q}^{Dplr}, 0 \leq q < Q$.
From (\ref{Doplr_DDRG_rect_expsion}) it then follows that
\vspace{-2mm}
\begin{eqnarray}{\label{vector_notation_doppler}}
	\mathbf{y}_{q'} \hspace{-3mm} &=& \mathbf{\Tilde H}_{q',q'}\mathbf{{\Tilde x}}_{q'}+ \mathbf{z}_{q^{\prime}}^{Dplr}, \nonumber \\
	\mathbf{z}_{q^{\prime}}^{Dplr} &\Define& \sum_{\substack{q=0 \\ q \ne q'}}^{Q-1}\mathbf{\Tilde H}_{q,q'} \mathbf{{\Tilde x}}_{q}+\mathbf{w}_{q'}
\end{eqnarray}
where $\mathbf{\Tilde H}_{q,q^{\prime}} \in \mathbb{C}^{M\left(\frac{N}{Q}-G\right) \times M\left(\frac{N}{Q}-G\right)}$ has $\Tilde h_{q,q^{\prime}}[\Tilde k,\Tilde l,k,l]$ as its element in its $ ((\Tilde k-q^{\prime}(N/Q))+\Tilde l ((N/Q)-G)+1$)-th  row and $ (( k-q(N/Q))+ l ((N/Q)-G)+1$)-th column. Since $w(t)$ is AWGN with power spectral density $N_0$, from (\ref{w_k_l_Doplr}) it follows that $w[\Tilde k,\Tilde l] \sim i.i.d. \hspace{2mm} \mathcal{CN}(0,MNN_{0})$, $((\Tilde k,\Tilde l) \in \mathcal{R}_{q^{\prime}}^{Dplr}, 0 \leq q^{\prime} < Q)$, and ${\Tilde x}_{q^{\prime}}[k,l] \sim i.i.d. \hspace{2mm} \mathcal{CN}(0,E_{T})$. From (\ref{vector_notation_doppler}) it then follows that the SE achieved by the $q^{\prime}$-th UT is given by \cite{Tse_Viswanath}
\begin{eqnarray}{\label{SSE_Doplr_final}}
	R_{q'}^{Dplr} &=& \frac{1}{MN} \log_{2} \left|\mathbf{I}+E_{T} \mathbf{\Tilde H}_{q',q'}^{H} \mathbf{K}_{{q'}}^{Dplr^{-1}} \mathbf{\Tilde H}_{q',q'}\right|
\end{eqnarray}
where $\mathbf{K}_{q^{\prime}}^{Dplr}  \hspace{-1mm} \Define \hspace{-1mm} \mathbb{E}\left[\mathbf{z}_{q^{\prime}}^{Dplr}\mathbf{z}_{q^{\prime}}^{Dplr^{H}}\right] = E_T  \hspace{-2mm} \sum\limits_{q=0, q \ne q'}^{Q-1}  \hspace{-2mm} \mathbf{\Tilde H}_{q,q'} \mathbf{\Tilde H}_{q,q'}^H   +  MNN_0 \mathbf{I}$ is the covariance matrix of $\mathbf{z}_{q^{\prime}}^{Dplr}$. \\
\subsubsection{Analysis of the sum SE achieved by Delay domain GB based MA method with rectangular transmit and receive pulses} 
From (\ref{Set_R_Dly,q}), it follows that the TF domain symbols at the $q$-th UT are given by
\vspace{-3mm}
\begin{eqnarray}
\label{newInvSFFT_Dly}
X_q[n,m] & = & \frac{1}{M N} \sum_{(k,l) \in {\mathcal R}_{q}^{Dly} }  \, {\Tilde x}_q[k,l] \, e^{-j 2 \pi {\Big (}   \frac{ml}{M}  - \frac{n k}{N} {\Big )}} \nonumber \\
&  \hspace{-5mm} \mya &  \hspace{-3mm}  \frac{1}{MN} \hspace{-3mm}\sum_{l=q\cdot\frac{M}{Q}}^{\left(\frac{M}{Q}-G-1\right)+q\cdot\frac{M}{Q}}\sum_{k=0}^{N-1} {\Tilde x}_{q}[k,l]e^{-j2\pi \left(\frac{ml}{M}-\frac{nk}{N}\right)} \nonumber \\
&\hspace{-26mm}  & \hspace{-16mm} n=0, \cdots, N-1, m=0,\cdots, M-1.
\end{eqnarray} 
where step $(a)$ follows from the fact that ${\Tilde x}_{q}[k,l]=0$ for $(k,l) \notin \mathcal{R}_{q}^{Dly}$. The transmitted TD signal, $s_{q}(t)$, of the $q$-th UT is then given by (\ref{Tx_signal}).

For the delay domain GB based MA method with rectangular transmit and receive pulses, the received TF domain symbols $Y[\Tilde n,\Tilde m]$ at the BS are given by Lemma \ref{lemma3_Interlvd_TF} (with $X_q[n,m]$ given by (\ref{newInvSFFT_Dly})).
The DD domain symbols received in the DDREs in the set $\mathcal{R}_{q^{\prime}}^{Dly}$ are used to demodulate the information symbols transmitted by the $q^{\prime}$-th UT. The following theorem gives the expression of the DD domain symbols received in $\mathcal{R}_{q^{\prime}}^{Dly}$ in terms of the DD domain information symbols transmitted by all UTs.
\begin{theorem}
    \label{Lemma5_Delay_DD}
    For the $q^{\prime}$-th UT ($0 \leq q^{\prime} < Q$), the DD domain symbols $y_{q^{\prime}}[\Tilde{k},\Tilde{l}]$ received at the BS for $(\Tilde{k},\Tilde{l}) \in \mathcal{R}_{q^{\prime}}^{Dly}$ are given by
    \begin{eqnarray}
			\label{Dly_DDRG_rect_expsion}
			\nonumber  y_{q^{\prime}}[\Tilde{k},\Tilde{l}] &=& \underbrace{\sum_{k=0}^{N-1}\sum_{l=q'\cdot \frac{M}{Q}}^{\left(\frac{M}{Q}-G-1\right)+q'\cdot \frac{M}{Q}} {\Tilde x}_{q'}[k,l] \hat{h}_{q',q'}[\Tilde{k},\Tilde{l},k,l]}_{\text{Useful term}} \\
			& & \hspace{-18mm} + \underbrace{\sum_{\substack{q=0 \\ q \ne q'}}^{Q-1} \sum_{k=0}^{N-1}\sum_{l=q\cdot \frac{M}{Q}}^{\left(\frac{M}{Q}-G-1\right)+q\cdot \frac{M}{Q}}\hspace{-5mm} {\Tilde x}_{q}[k,l] \hat{h}_{q,q'}[\Tilde{k},\Tilde{l},k,l]}_{\text{MUI term}} + \underbrace{w[\Tilde{k},\Tilde{l}]}_{\text{AWGN}}
	\end{eqnarray}
		where $\hat{h}_{q,q'}[\Tilde{k},\Tilde{l},k,l]$ and $w[\Tilde{k},\Tilde{l}]$ are defined in (\ref{Dly_h_q_q_rect}) and (\ref{w_k_l_Dly}) respectively (in Appendix \ref{proof_of_Dly_DDRG_lemma}) and $(\Tilde k,\Tilde l) \in  \mathcal{R}_{q^{\prime}}^{Dly}$.
\end{theorem}
	\begin{IEEEproof}
		See Appendix \ref{proof_of_Dly_DDRG_lemma}.
	\end{IEEEproof}

Let the DD signal $y_{q^{\prime}}[\Tilde k,\Tilde l]$, $((\Tilde k,\Tilde l) \in \mathcal{R}_{q^{\prime}}^{Dly}, 0 \leq q^{\prime} < Q)$, be arranged into the
vector $\mathbf{y}_{q^{\prime}} \in \mathbb{C}^{N\left(\frac{M}{Q}-G\right)\times1}$ such that
the  $(\Tilde k ((M/Q)-G)+(\Tilde l-q^{\prime}(M/Q))+1)$-th element of $\mathbf{y}_{q^{\prime}}$ is $y_{q^{\prime}}[\Tilde k,\Tilde l]$.
Similarly let $\mathbf{w}_{q^{\prime}} \in  \mathbb{C}^{N\left(\frac{M}{Q}-G\right)\times1}$ denote the vector of AWGN samples $w[\Tilde k,\Tilde l]$, $((\Tilde k,\Tilde l) \in \mathcal{R}_{q^{\prime}}^{Dly}, 0 \leq q^{\prime} < Q)$. Also let $\mathbf{{\Tilde x}}_{q} \in \mathbb{C}^{N\left(\frac{M}{Q}-G\right)\times1}$ denote the vector of DD domain information symbols transmitted by the $q$-th UT, where the $(k ((M/Q)-G)+( l-q(M/Q))+1)$-th element of this vector is ${\Tilde x}_{q}[k, l]$, where $( k, l) \in \mathcal{R}_{q}^{Dly}, 0 \leq q < Q$.
From (\ref{Dly_DDRG_rect_expsion}) it then follows that
\begin{eqnarray}{\label{vector_notation_delay}}
	\mathbf{y}_{q'} \hspace{-3mm} &=& \mathbf{\widehat{H}}_{q',q'}\mathbf{{\Tilde x}}_{q'}+ \mathbf{z}_{q^{\prime}}^{Dly}, \nonumber \\
	\mathbf{z}_{q^{\prime}}^{Dly} &\Define& \sum_{\substack{q=0 \\ q \ne q'}}^{Q-1}\mathbf{\widehat{H}}_{q,q'} \mathbf{{\Tilde x}}_{q}+\mathbf{w}_{q'}
\end{eqnarray} 
where $\mathbf{\widehat{H}}_{q,q^{\prime}} \in \mathbb{C}^{N\left(\frac{M}{Q}-G\right) \times N\left(\frac{M}{Q}-G\right)}$ has $\hat{h}_{q,q^{\prime}}[\Tilde k,\Tilde l,k,l]$ as its element in the $(\Tilde k ((M/Q)-G)+(\Tilde l-q^{\prime}(M/Q))+1)$-th row and ($k ((M/Q)-G)+( l-q(M/Q))+1$)-th column. Since $w(t)$ is AWGN having power spectral density $N_0$, from (\ref{w_k_l_Dly}) it follows that $w[\Tilde k,\Tilde l] \sim i.i.d. \hspace{2mm} \mathcal{CN}(0,MNN_{0})$, $((\Tilde k,\Tilde l) \in \mathcal{R}_{q^{\prime}}^{Dly}, 0 \leq q^{\prime} < Q)$. Also, let ${\Tilde x}_{q^{\prime}}[k,l] \sim i.i.d. \hspace{2mm} \mathcal{CN}(0,E_{T})$.
From (\ref{vector_notation_delay}) it then follows that the SE achieved by the $q^{\prime}$-th UT is given by \cite{Tse_Viswanath}
\begin{eqnarray}{\label{SSE_Dly_final}}
	R_{q'}^{Dly} &=& \frac{1}{MN} \log_2 \left|\mathbf{I}+E_{T} \mathbf{\widehat{H}}_{q',q'}^{H} \mathbf{K}_{{q'}}^{Dly^{-1}} \mathbf{\widehat{H}}_{q',q'}\right|
\end{eqnarray}
where $\mathbf{K}_{q^{\prime}}^{Dly} \hspace{-1mm}\Define\hspace{-1mm} \mathbb{E}\left[\mathbf{z}_{q^{\prime}}^{Dly}\mathbf{z}_{q^{\prime}}^{Dly^{H}}\right]= E_T  \hspace{-2mm} \sum\limits_{q=0, q \ne q'}^{Q-1}  \hspace{-2mm} \mathbf{ \widehat{H}}_{q,q'} \mathbf{ \widehat{H}}_{q,q'}^H   +  MNN_0 \mathbf{I}$ is the covariance matrix of $\mathbf{z}_{q^{\prime}}^{Dly}$.



	\section{Numerical Results}
	\label{Num_simulations}
	In this section, we compare the average sum SE of IDDMA, ITFMA and the GB based MA methods in an OTFS based communication system where \emph{practical rectangular transmit and receive pulses} are used. For the simulations, we consider $\Delta f = 1/T = 15$ KHz, $M=36, N=18$. For each UT the Extended Typical Urban (ETU 300) channel model standardized by 3GPP \cite{3GPPmdl} is considered, where the maximum Doppler shift is $\nu_{max} = 300$ Hz. The delay profile $\{  \tau_{q,i} \}_{i=1}^{p_q=9}$ is $ [0 \,,\, 50 \,,\,  120 \,,\, 200  \,,\, 230  \,,\, 500  \,,\, 1600 \,,\,$ $2300 \,,\, 5000]$ ns and the corresponding power profile $\{ {\mathbb E}[ | h_{q,i} |^2 ] \}_{i=1}^{p_q = 9}$ is $[-1 \,,\, -1 \,,\, -1 \,,\, 0 \,,\, 0 \,,\, 0  \,,\, -3 \,,\, -5 \,,\, -7]$ dB. We further normalize this power profile such that $\sum\limits_{i=1}^{9} \, {\mathbb E}[| h_{q,i} |^2] \, = \, 1 $. The channel gains $h_{q,i}$ are independent Rayleigh faded random variables. For the $i$-th channel path, the Doppler shift is modeled as $\nu_{q,i} = \nu_{max}\cos\left(\theta_{q,i}\right)$ where $\theta_{q,i}$ ($0 \leq q <Q, i =1,2,\cdots,p_{q}$) are independent and uniformly distributed in $[0, 2\pi)$. \\
	For plotting the average sum SE of the IDDMA method with rectangular pulses, we use the sum rate expression $\sum_{q^{\prime}=0}^{Q-1}R_{q^{\prime}}$, where $R_{q^{\prime}}$ is given by (\ref{SSE_proposed}). Similarly for the sum SE of Doppler and delay domain GB based MA methods with rectangular pulses we use the sum rate expression $\sum_{q^{\prime}=0}^{Q-1}R_{q^{\prime}}^{Dplr/Dly}$, where the expression for $R_{q^{\prime}}^{Dplr}$ and $R_{q^{\prime}}^{Dly}$ are given by (\ref{SSE_Doplr_final}) and (\ref{SSE_Dly_final}) respectively.
The sum SE of the ITFMA method with rectangular pulses is given by $\sum_{q'=0}^{Q-1} R_{q'}^{\mbox{\tiny{Int}}}$ where $R_{q'}^{\mbox{\tiny{Int}}}$ is given by (\ref{SSE_intrlvd_final}).

	In the following, the ratio of the transmitted signal power from a UT (i.e., $\mathbb{E}\left[\int|s_{q}(t)|^{2}dt/(NT)\right]$) to the receiver AWGN power in the total bandwidth $M\Delta f$ is referred to as the signal-to-noise ratio (SNR).  For the IDDMA method, from the expressions derived in Section \ref{iddmasectionref} it follows that the average received SNR is $\rho/Q^{2}$,  where $\rho \Define E_{T}/(MNN_{0})$. For the ITFMA method, from the expressions derived in Section \ref{itfmasectionref} it follows that the average received SNR is $\rho$. For the Doppler GB based MA method, from the expressions derived in Section \ref{gbmasectionref} it follows that the average received SNR is $\rho\left(\frac{1}{Q}-\frac{G}{N}\right)$, where $G$ is the guard band size. For the delay GB based MA method, from the expressions derived in Section \ref{gbmasectionref} it follows that the average received SNR is  $\rho\left(\frac{1}{Q}-\frac{G}{M}\right)$.
	
	\begin{figure}[!t]
		\centering
		\includegraphics[width=\linewidth]{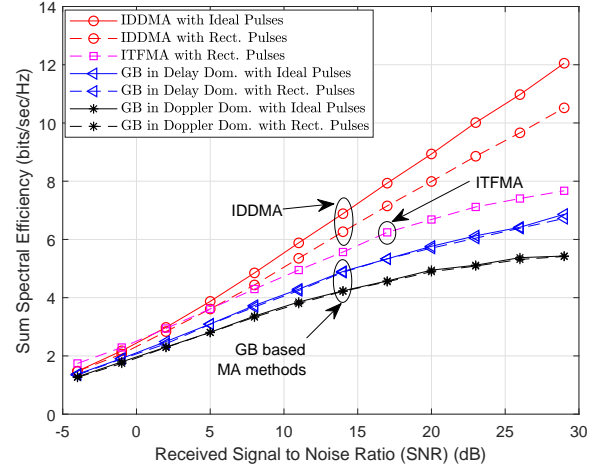}  
		\caption{Sum Spectral Efficiency vs. SNR ($M = 36,N = 18$, $Q = 6, g_1 = g_3 = 3,g_{2} = g_4 = 2, \nu_{max} = 300$ Hz).}
		\label{fig:SSE_Q6}
		\vspace{-5mm}
	\end{figure}
	In Fig. \ref{fig:SSE_Q6}, we plot the sum SE versus the SNR for $Q=6$ UTs and $\nu_{max} = 300$ Hz. For the IDDMA method we consider $g_{1}=3, g_{2}=2$ and for the ITFMA method we consider $g_3 = 3, g_4 = 2$. For the GB based methods we choose the guard band size $G$ such that it maximizes the sum SE. With rectangular pulses, it is observed that for a given practical desired per-UT SE (e.g., per-UT SE more than $0.5$ bits/sec/Hz) the SNR required by the IDDMA method is smaller than that required by the ITFMA  method which is in turn smaller than that required by the GB based methods. This is because, in order to reduce MUI, the GB based MA methods use guard bands which do not carry any information. The required SNR is smaller for the IDDMA method when compared to ITFMA because, the allocation of TFREs is contiguous in IDDMA whereas it is interleaved  in ITFMA (see Fig.~\ref{fig0} and Fig.~\ref{fig3_int}). Due to interleaved TFRE allocation in ITFMA, \emph{each} TF symbol of a particular UT experiences MUI along the time-domain (due to channel delay spread) and along the frequency domain (due to Doppler shift) from TF symbols transmitted by other UTs in neighbouring TFREs. However, due to the contiguous TFRE allocation in IDDMA, \emph{only} the TF symbols transmitted on the TFREs located at the boundary of the allocated TFRE region of a UT get affected by interference from TF symbols transmitted on adjacent TFREs allocated to other UTs whereas, the TF symbols transmitted on TFREs in the interior of the allocated TFRE region of that UT are not as much affected by transmissions from other UTs. Therefore, the amount of MUI experienced in IDDMA is lesser than that in ITFMA.  
	
	In Fig. \ref{fig:SSE_Q6} we have also plotted the sum SE achieved by the IDDMA and the GB based methods with ideal pulses. For the IDDMA method, the sum SE achieved for both ideal and rectangular pulses is the same at low SNR. However at high SNR, the SE achieved with rectangular pulses degrades slightly due to MUI, ICI and ISI. This is because with rectangular pulses, at high SNR the effective noise and interference is dominated by interference (MUI, ICI and ISI), whose power increases with increasing SNR.
	
	 In Fig. \ref{fig:SSE_Q6}, we also observe that the sum SE achieved by the GB based MA methods is the same for both ideal and rectangular pulses. This is because, even with ideal pulses there is MUI due to non-zero Doppler and delay spread of the wireless channel.\footnote{\footnotesize{In the ETU300 channel model, the maximum delay spread is $5 \mu s$. With $M = 36$ and $\Delta f = 15$ KHz, the delay domain resolution $T/M = 1.85 \mu s$, i.e., the maximum delay spread is roughly $\lceil 5/1.85 \rceil = 3$ DDREs along the delay domain. With $\nu_{max} = 300$ Hz and $N=18$, i.e., a Doppler domain resolution of $1/(NT) = \Delta f/N = 833$ Hz, the maximum Doppler spread is one DDRE along the Doppler domain.}} The additional MUI due to the use of rectangular pulses is not as significant as the MUI due to multi-path delay and Doppler spread.
	\begin{figure}[ht]
		\centering
		\includegraphics[width=\linewidth]{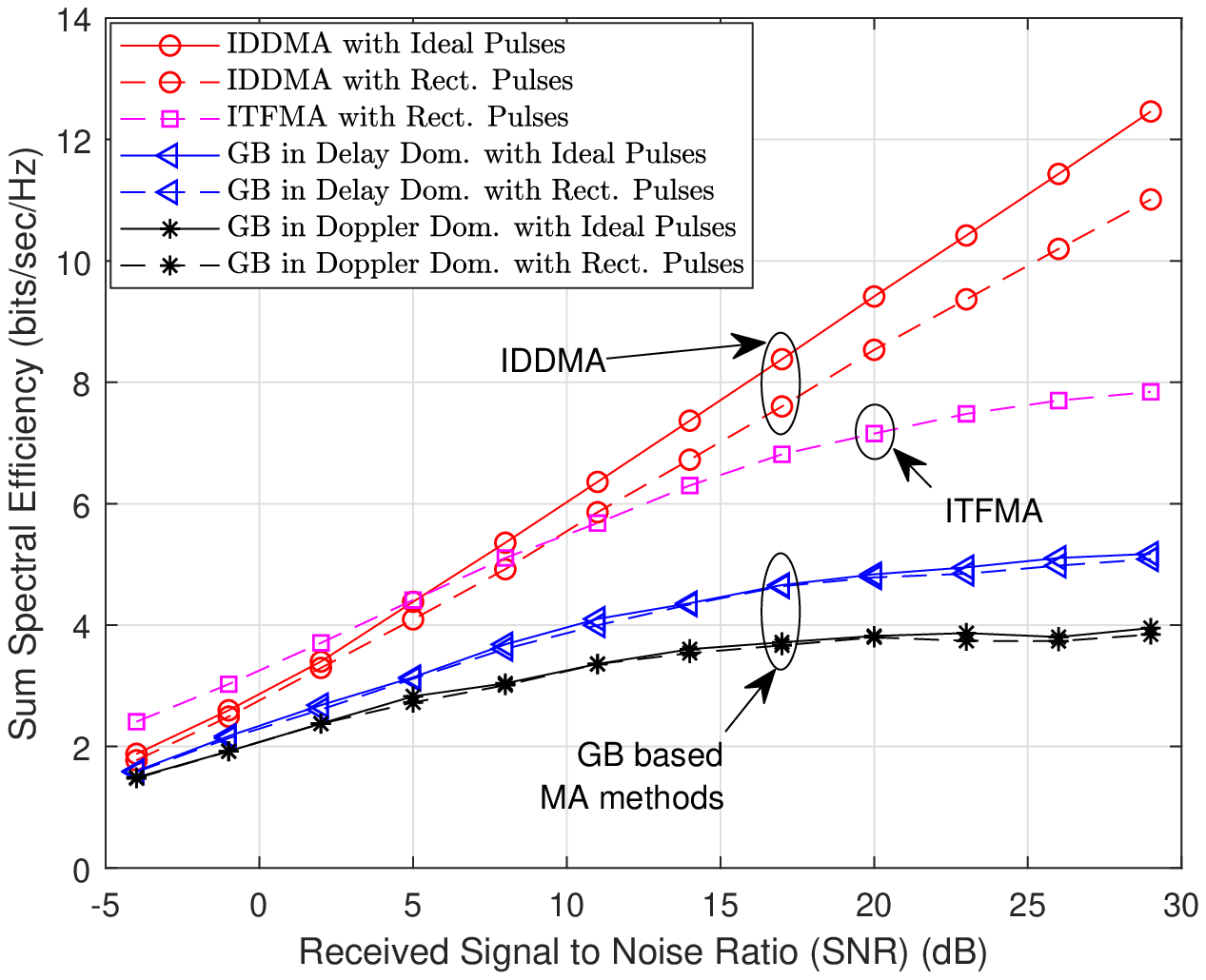}  
		\caption{Sum Spectral Efficiency vs. SNR ($M = 36,N = 18$, $Q = 9, g_1 = g_3 = 3, g_2 = g_4 = 3,\nu_{max} = 300$ Hz).}
		\label{fig:SSE_Q9}
		\vspace{-2mm}
	\end{figure} 
	\begin{figure}[ht]
		\centering
		\includegraphics[width=\linewidth]{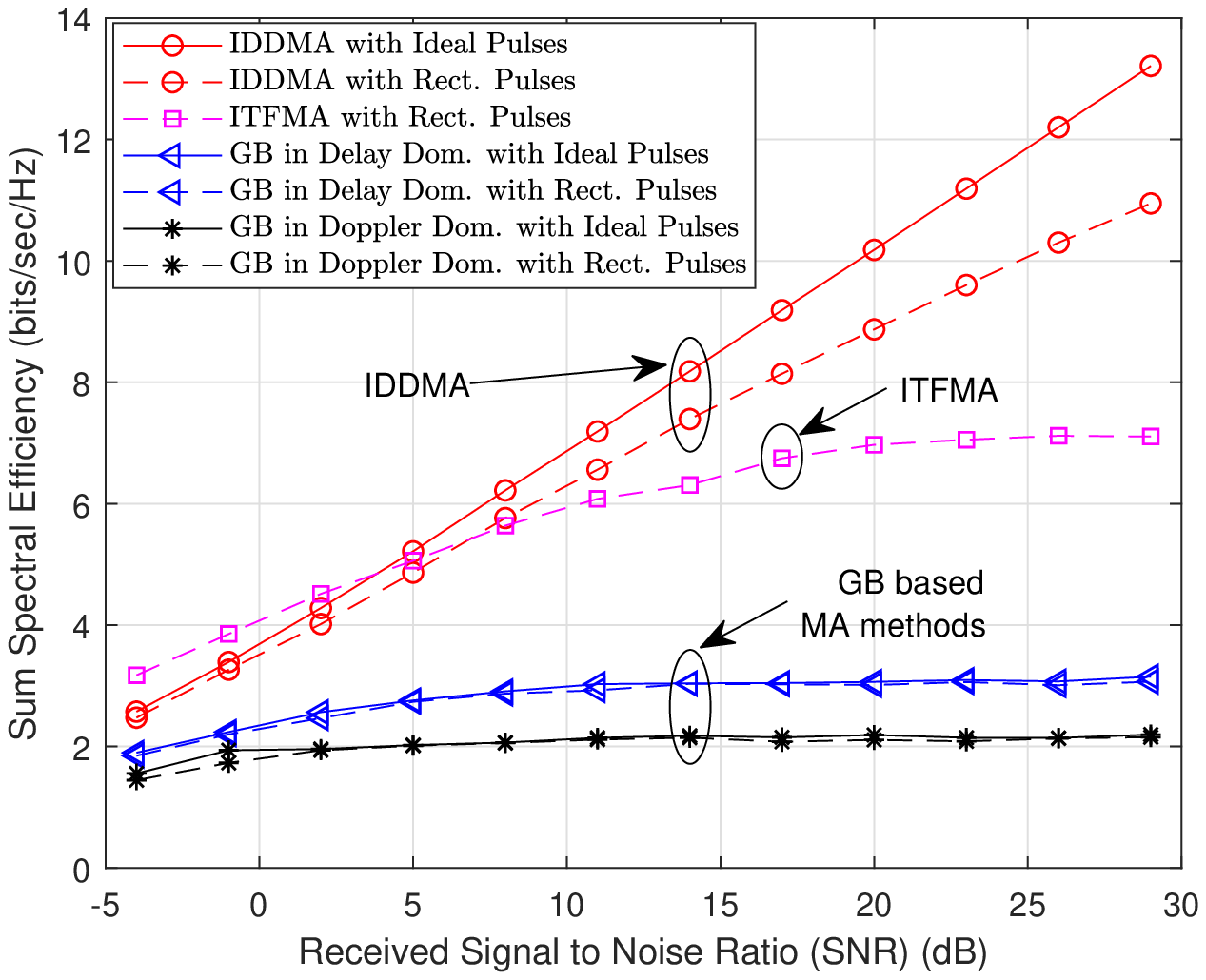}  
		\caption{Sum Spectral Efficiency vs. SNR ($M = 36,N = 18, Q = 18,g_1 = g_3 = 6, g_2 = g_4 = 3, \nu_{max} = 300$ Hz).}
		\label{fig:SSE_Q18}
		\vspace{-3mm}
	\end{figure}
	
	In Fig. \ref{fig:SSE_Q9} and Fig. \ref{fig:SSE_Q18}, we plot the sum SE versus SNR with $Q=9$ UTs and $Q=18$ UTs respectively and $\nu_{max}=300$ Hz.
	Similar observations as that from Fig. \ref{fig:SSE_Q6} can also be made in Fig. \ref{fig:SSE_Q9} and Fig. \ref{fig:SSE_Q18}. 
	
\begin{figure}[ht]
		\centering
		\includegraphics[width=\linewidth]{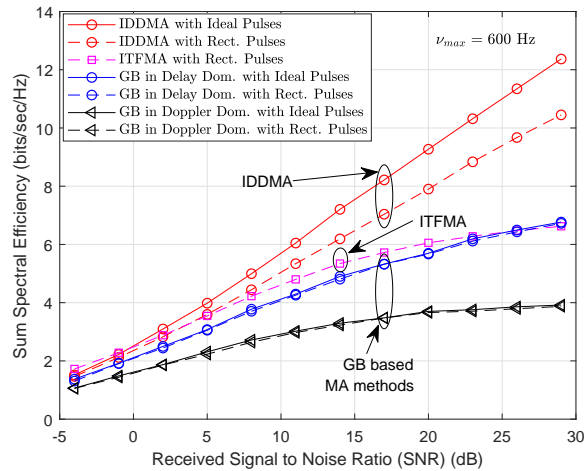}  
		\caption{Sum Spectral Efficiency vs. SNR ($M = 36,N = 18, Q = 6, g_1 = g_3 = 3,g_{2} = g_4 = 2, \nu_{max} = 600$ Hz).}
		\label{fig:SSE_Q6_nu600}
	\end{figure}

\begin{figure}[ht]
		\centering
		\includegraphics[width=\linewidth]{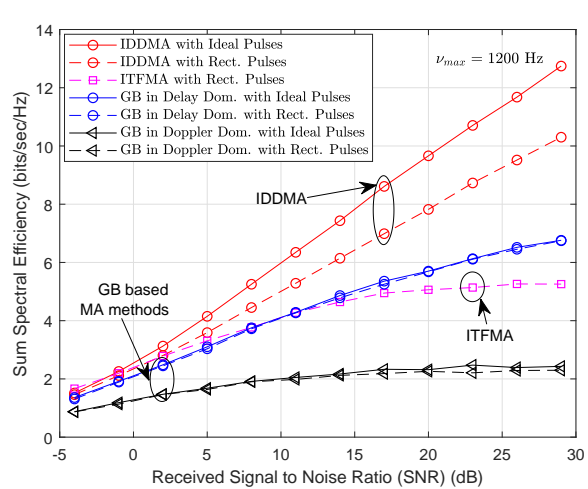}  
		\caption{Sum Spectral Efficiency vs. SNR ($M = 36,N = 18, Q = 6, g_1 = g_3 = 3,g_{2} = g_4 = 2, \nu_{max} = 1200$ Hz).}
		\label{fig:SSE_Q6_nu1200}
	\end{figure}

	
	In Fig. \ref{fig:SSE_Q6_nu600} and Fig. \ref{fig:SSE_Q6_nu1200}, we plot the sum SE versus SNR for $\nu_{max}=600$ Hz and $\nu_{max}=1200$ Hz respectively, with $Q=6$ UTs ($g_1 = g_3 = 3, g_2=g_4 = 2$). We observe that for both $\nu_{max}= 600, 1200$ Hz, for practical SNR (at which the per-UT SE is $0.5$ bits/sec/Hz or higher), the IDDMA method achieves a better sum SE performance when compared to the ITFMA and the GB based MA methods. Comparing Fig. \ref{fig:SSE_Q6_nu600} and Fig. \ref{fig:SSE_Q6_nu1200}, it is observed that the sum SE achieved by the Doppler domain GB based MA method decreases significantly with increasing $\nu_{max}$. This is because, with increasing $\nu_{max}$ the Doppler domain GBs are not sufficient to reduce interference between different UTs along the Doppler domain due to the increased multi-path Doppler shift. Also the sum SE achieved by the delay domain GB based MA method is almost invariant to increase in $\nu_{max}$, as the UTs are separated by GBs along the delay domain and therefore the interference between UTs is only due to the multi-path delay spread which does not change with increasing $\nu_{max}$.

\begin{figure}[ht]
		\centering
		\includegraphics[width=\linewidth]{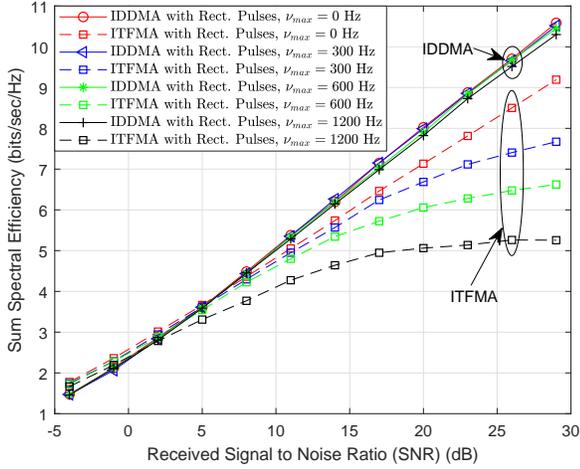}  
		\caption{Sum Spectral Efficiency vs. SNR ($M = 36,N = 18, Q = 6, g_1 = g_3 = 3,g_{2} = g_4 = 2$).}
		\label{fig:SSE_Q6_allnu_proposed}
		\vspace{-5mm}
	\end{figure}

	By comparing Fig. \ref{fig:SSE_Q6_nu600} and Fig. \ref{fig:SSE_Q6_nu1200}, it is also observed that
	the sum SE achieved by the IDDMA method with rectangular pulses is almost the same for both $\nu_{max}= 600, 1200$ Hz, whereas the sum SE for the ITFMA method \emph{decreases} when $\nu_{max}$ is increased from $600$ Hz to $1200$ Hz. In fact, for $\nu_{max} = 1200$ Hz, the sum SE performance of ITFMA is inferior to that of the delay domain GB based method at high SNR.
	To understand this better, in Fig. \ref{fig:SSE_Q6_allnu_proposed}, we plot the sum SE achieved by the IDDMA and the ITFMA methods (with rectangular pulses) as a function of increasing SNR for $\nu_{max}=0,300,600,1200$ Hz, with $Q=6$ UTs ($g_{1}=g_3=3,  g_{2}=g_4=2$).
	In Fig.~\ref{fig:SSE_Q6_allnu_proposed}, we observe that with rectangular pulses the sum SE performance of the IDDMA method is \emph{almost invariant} of the maximum Doppler shift $\nu_{max}$,
	whereas for the ITFMA method the sum SE performance \emph{decreases monotonically} with increasing $\nu_{max}$. This is due to the interleaved TFRE allocation in ITFMA where \emph{each} TF symbol of a particular UT experiences MUI from TF symbols transmitted by other UTs in neighbouring TFREs. In fact due to the interleaved TFRE allocation, the sum SE performance
	of ITFMA is inferior to that of IDDMA even when $\nu_{max} = 0$ Hz (i.e., no mobility scenario). This is because, in ITFMA even in the no mobility scenario, each TF symbol of a UT still experiences MUI along the time-domain due to the delay-spread of the multi-path channel of other UTs.      
	\begin{figure}[ht]
		\centering
		\includegraphics[width=\linewidth]{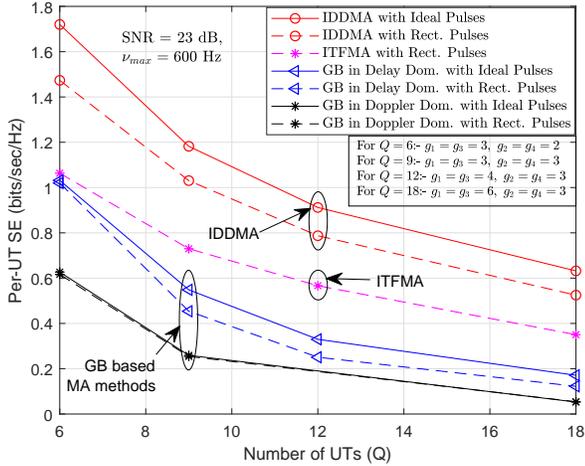}  
		\caption{Per-UT Spectral Efficiency vs. $Q$ ($M = 36,N = 18, \nu_{max} = 600$ Hz).}
		\label{fig:perUTSE_2}
		\vspace{-3mm}
	\end{figure} 
	
	In Fig.~\ref{fig:perUTSE_2} we plot the per-UT SE achieved by the MA methods versus the number of UTs, for a fixed SNR of $23$ dB. For $Q=6,9, 18$, the values of $g_1,g_2,g_3,g_4$ are same as that considered in Fig.  \ref{fig:SSE_Q6}, Fig. \ref{fig:SSE_Q9} and Fig. \ref{fig:SSE_Q18}. In Fig.~\ref{fig:perUTSE_2}, for $Q=12$ we have $g_1=g_3 = 4, g_2=g_4=3$. It is observed that for all MA methods, the per-UT SE decreases
	with increasing number of UTs. This decrease is primarily due to the decrease in the number of DDREs allocated to each UT, with increasing number of UTs.
	It is observed that the per-UT SE achieved by the IDDMA method is better than that achieved by the ITFMA and the GB based MA methods.  
	For the IDDMA method, we further observe that the gap between the per-UT SE achieved with rectangular pulses and that achieved with ideal pulses is small for all $Q=6,9,12, 18$. 

	\section{Conclusion}
	\label{sec_8}
	In this paper, we have derived non-trivial expressions of the received DD domain symbols
	for the IDDMA, ITFMA and the GB based MA methods, when practical rectangular pulses
	are used instead of ideal pulses. We also derive the expression for the sum SE achieved by these methods. Through simulations,
	it is observed that for practical values of the received SNR, with rectangular pulses the IDDMA method achieves higher sum SE than that achieved
	by the ITFMA and the GB based methods. We also observe that the sum SE achieved by the IDDMA method
	is almost invariant of the maximum Doppler shift whereas the sum SE achieved by the ITFMA method decreases monotonically with increasing maximum Doppler shift.
	This shows the robustness of the IDDMA method towards mobility induced Doppler shift when compared to ITFMA and the GB based MA methods, when practical rectangular pulses
	are used.       
	
	
	%

	\begin{appendices}
		\section{Proof of Lemma \ref{lemma_1}}
    	\label{proof_of_lemma_1}
    	 After substituting (\ref{rect_pulses}) and (\ref{rx_signal_r(t)}) in (\ref{Yqprime}), the received TF symbols $Y_{q^{\prime}}[\Tilde{n},\Tilde{m}]$ are given by 
    	 \vspace{-3mm}
	\begin{eqnarray}
		\label{Yprime_rect_1}
			\nonumber Y_{q'}[\Tilde n,\Tilde m]  \hspace{-3mm} &=& \hspace{-3mm} \frac{1}{\sqrt{T}}\sum_{q=0}^{Q-1}\sum_{i=1}^{p_q}\sum_{m=0}^{\frac{M}{g_{1}}-1}\sum_{n=0}^{\frac{N}{g_{2}}-1} h_{q,i}X_{q}[n,m] e^{-j2\pi \nu_{q,i} \tau_{q,i}} \\
			\nonumber & & \hspace{-5mm} e^{-j2\pi \big(m+\frac{M}{g_{1}}\lfloor q / g_{2}\rfloor\big)\Delta f \tau_{q,i}} F[n, \Tilde n, m, \Tilde m, q,q^{\prime}] \\
	         & & \qquad\qquad\qquad\qquad\qquad+W_{q^{\prime}}[\Tilde n,\Tilde m]
	\end{eqnarray}
	where, \vspace{-5mm}
	\begin{eqnarray} 
	        \label{integral_rect}
			 F[n, \Tilde n, m, \Tilde m, q,q^{\prime}] \hspace{-3mm} & \Define & \hspace{-3mm} \int_{\left(\Tilde n+\frac{N}{g_{2}}(q')_{g_{2}}\right)T}^{\left(\Tilde n+1+\frac{N}{g_{2}}(q')_{g_{2}}\right)T} e^{j2\pi \xi \frac{t}{T}}  \nonumber \\ 
			& & g_{tx}\Big(t-\tau_{q,i}-\big(n+\frac{N}{g_{2}}(q)_{g_{2}}\big)T\Big) dt  \nonumber \\
			\hspace{-5mm}\xi \hspace{-3mm} &\Define& \hspace{-3mm}  \big(m-\Tilde m \hspace{-1mm}+\hspace{-1mm}\frac{M}{g_{1}}\left(\left\lfloor \frac{q}{g_{2}}\right\rfloor\hspace{-1mm}-\hspace{-1mm}\left\lfloor \frac{q^{\prime}}{g_{2}}\right\rfloor\right)+T\nu_{q,i}\big) \nonumber \\
			\hspace{-7mm} & & \hspace{-7mm}  q, q^{\prime} = 0,1,\cdots,Q-1.
	\end{eqnarray}
	\begin{eqnarray} \label{noise_rect_11}
		W_{q'}[\Tilde n,\Tilde m] \hspace{-3mm} &\Define& \hspace{-3mm}  \frac{1}{\sqrt{T}}\int_{\left(\Tilde n+\frac{N}{g_{2}}(q')_{g_{2}}\right)T}^{\left(\Tilde n+1+\frac{N}{g_{2}}(q')_{g_{2}}\right)T} \hspace{-7mm}w(t) e^{-j2\pi \big(\Tilde m+\frac{M}{g_{1}}\left\lfloor \frac{q^{\prime}}{g_{2}} \right\rfloor\big) \Delta f t} \hspace{-1mm} dt. \nonumber \\
	\end{eqnarray}
	After simplifying the integral in (\ref{integral_rect}), $F[n, \Tilde n, m, \Tilde m, q,q^{\prime}]$,  is given by
		\begin{eqnarray} \label{intergral_rect_sol_1}
			\nonumber  F[n, \Tilde n, m, \Tilde m, q,q^{\prime}] \hspace{-3mm}  &=& \hspace{-3mm} e^{j2\pi \big(m-\Tilde m+\frac{M}{g_{1}}\big(\lfloor q / g_{2}\rfloor-\lfloor q' / g_{2}\rfloor\big)\big)\Delta f \tau_{q,i}} \nonumber \\
			& & \hspace{-24mm}  e^{j2\pi \nu_{q,i} \Big(\tau_{q,i}+\big(n+\frac{N}{g_{2}}(q)_{g_{2}}\big)T\Big)} \int_{t'_{l}}^{t'_{u}} \hspace{-2mm} g_{tx}(t) e^{j2\pi \xi \frac{t}{T}}  dt
		\end{eqnarray}
		where,
		\begin{eqnarray}
		\label{tludefeqn}
			t'_{l} & \Define & \left(\Tilde n-n+\frac{N}{g_{2}}\big((q')_{g_{2}}-(q)_{g_{2}}\big)\right)T-\tau_{q,i}  \nonumber \\
			t'_{u} & \Define & \left(\Tilde n-n+1+\frac{N}{g_{2}}\big((q')_{g_{2}}-(q)_{g_{2}}\big)\right)T-\tau_{q,i}.
		\end{eqnarray}
		From the above definitions of $t'_{l}$ and $t'_{u}$, and the fact that $g_{tx}(t)$ is non-zero only for $t \in [0 \,,\, T]$, it is clear that the integral in the R.H.S. of (\ref{intergral_rect_sol_1}) is
		zero when $\left\vert  (q')_{g_{2}}-(q)_{g_{2}} \right\vert \geq 2$. Therefore, next we compute the lower and upper limits (i.e., $t'_l$ and $t'_u$) for the integral in the R.H.S. of (\ref{intergral_rect_sol_1}), for three different cases, i.e., when
		$\left((q')_{g_{2}}-(q)_{g_{2}} \right) = 0,\pm 1$. 	
		\subsection{\underline{When $(q')_{g_2} - (q)_{g_2} = 0$}:-}
		We know that $0\le q \le Q-1$, where $Q=g_1g_2$. For $(q')_{g_2}=(q)_{g_2}$, from (\ref{tludefeqn}) we have
		\begin{eqnarray}
		\label{tlucasea}
		t'_{l} & =  & \left(\Tilde n-n\right)T-\tau_{q,i}, \nonumber \\ 
		t'_{u} & =  & \left(\Tilde n-n+1\right)T-\tau_{q,i}.
		\end{eqnarray}
		From (\ref{tlucasea}) we note that for this scenario, $t'_{u} = t'_{l} + T$. However, since $0 \leq \tau_{q,i} < T$ (see (\ref{max_tau_nu})) and $g_{tx}(t)$ is non-zero only for $t \in [0 \,,\, T)$ (see (\ref{rect_pulses})), it follows that the integral in the R.H.S.
		of (\ref{intergral_rect_sol_1}) is non-zero if and only if ${\Tilde n} - n = 1,0$.
		\subsubsection*{ A.1) \underline{For $n=\Tilde n$}} Under this condition, from (\ref{tlucasea}) we have $t'_{l} = -\tau_{q,i}$ and  $t'_{u}=T-\tau_{q,i}$.
		Since $g_{tx}(t)$ is non-zero only in the interval $t \in [0 \,,\, T)$ it suffices to have $t'_{l} = 0$.
		\subsubsection*{A.2) \underline{For $\Tilde n=n+1$}} Under this condition, from (\ref{tlucasea}) we have $t'_{l} = T-\tau_{q,i}$ and $t'_{u}=2T-\tau_{q,i}$.
		Since $0 \leq \tau_{q,i} < T$, it suffices to have $t'_{u}= T$.
		\subsection{\underline{When  $(q')_{g_2}-(q)_{g_2}=1$}:-} For this condition, from (\ref{tludefeqn}) we have $t'_{l} =  (\Tilde n-n+(N/g_{2}))T-\tau_{q,i}$ and $t'_{u} = (\Tilde n-n+1+(N/g_{2}))T-\tau_{q,i}$.
		We note that $n, {\Tilde n} \in \{ 0, 1,\cdots, N/g_2 - 1 \}$. For the integral in the R.H.S. of (\ref{intergral_rect_sol_1}) to be non-zero, there must be some overlap between the intervals $[t'_l \,,\, t'_u)$ and
		$[0 \,,\, T)$. For $0 \leq n \leq N/g_2 - 2$, we see that $t'_l = (\Tilde n-n+(N/g_{2}))T-\tau_{q,i} \geq (2 +  \Tilde n)T - \tau_{q,i} $. Further, since ${\Tilde n} \geq 0$ and $\tau_{q,i} < T$ we have
		$t'_l \geq T$ and hence the integral in the R.H.S. of (\ref{intergral_rect_sol_1}) is zero for $0 \leq n \leq N/g_2 - 2$.  Hence, we consider $ { n} = N/g_2 - 1$, for which we have $t'_{l} =  (\Tilde n-n+(N/g_{2}))T-\tau_{q,i} = (\Tilde n + 1)T-\tau_{q,i}$ and $t'_u =  (\Tilde n-n+1+(N/g_{2}))T-\tau_{q,i} =  (\Tilde n + 2)T-\tau_{q,i}$. Clearly, for non-zero overlap between $[t'_l \,,\, t'_u)$ and
		$[0 \,,\, T)$, we must have ${\Tilde n} = 0$. 
		\subsubsection*{B.1) \underline{For $n=(N/g_{2})-1$ and $\Tilde n = 0$}} Under this condition, $t'_{l} = T-\tau_{q,i}$  and $t'_{u}=2T-\tau_{q,i}$.
		Since $g_{tx}(t)$ is non-zero only in the interval $[0 \,,\, T)$ and $0 \leq \tau_{q,i} < T$, it suffices to have $t'_u = T$. 
		
		\subsection{\underline{When  $(q')_{g_2}-(q)_{g_2}=-1$}:-} Under this condition, from (\ref{tludefeqn}) we have $t'_{l} =  (\Tilde n-n-(N/g_{2}))T-\tau_{q,i}$ and $t'_{u} = (\Tilde n-n+1-(N/g_{2}))T-\tau_{q,i}$.
		We know that $n, {\Tilde n} \in \{ 0, 1,\cdots, N/g_2 - 1 \}$. Therefore for any $n, {\Tilde n}$, $({\Tilde n} - n) \leq N/g_2 - 1$. Hence, $	t'_{u} = (\Tilde n-n+1-(N/g_{2}))T-\tau_{q,i} \leq - \tau_{q,i} \leq 0$ since
		$0 \leq \tau_{q,i} < T$ (see (\ref{max_tau_nu})).	 Therefore, under this condition the integral in the R.H.S. of (\ref{intergral_rect_sol_1}) is zero for any $n,{\Tilde n}$.
		
	After considering all the above determined conditions for the integral in the R.H.S. of (\ref{intergral_rect_sol_1}), the final expression for $F[n, \Tilde n, m, \Tilde m, q,q^{\prime}]$ is given by (\ref{integral_rect_solutn}) (see top of next page).   
	 \begin{figure*}[!t]
	 \vspace{-8mm}
\small
	\begin{eqnarray}
		\label{integral_rect_solutn}
		F[n, \Tilde n, m, \Tilde m, q,q^{\prime}] \hspace{-3mm} &=& \hspace{-3mm} \sqrt{T} e^{j2\pi \big(\xi-T\nu_{q,i}\big)\Delta f \tau_{q,i}}e^{j2\pi \nu_{q,i} \Big(\tau_{q,i}+\big(n+\frac{N}{g_{2}}(q)_{g_{2}}\big)T\Big)}  \hspace{-1mm} \left\{ \hspace{-1mm}  \delta\left[(q)_{g_{2}}\hspace{-1mm} - \hspace{-1mm} (q^{\prime})_{g_{2}}\right]\hspace{-1mm} \left\{ \hspace{-1mm}  \left[\frac{e^{j2\pi\xi \left(1-\frac{\tau_{q,i}}{T}\right)}-1}{j2\pi\xi }\right]\right.\right. \delta[n-\Tilde n] \nonumber \\
		 & & \hspace{-30mm} + e^{j2\pi\xi}\left.\frac{1-e^{-j2\pi\xi\frac{\tau_{q,i}}{T}}}{j2\pi\xi} \delta\left[n-(\Tilde n-1)\right]\right\} + e^{j2\pi\xi}\frac{1-e^{-j2\pi\xi\frac{\tau_{q,i}}{T}}}{j2\pi\xi}\delta[\Tilde n]\delta\left[(q)_{g_{2}}-\left((q^{\prime})_{g_{2}}-1\right)\right] \left.\delta\left[n-\left(\frac{N}{g_{2}}-1\right)\right]\right\}.
	\end{eqnarray}
	\normalsize
		\hrulefill
 	\vspace*{4pt}
    \end{figure*}

	After substituting the expression of $F[n, \Tilde n, m, \Tilde m, q,q^{\prime}]$ from (\ref{integral_rect_solutn}) in (\ref{Yprime_rect_1}), the received TF symbols $Y_{q^{\prime}}[\Tilde n, \Tilde m]$ are given by (\ref{Rx_TFRG_simplified_exp_rect}) where $H_{q,q^{\prime},1}[m,\Tilde n,\Tilde m], H_{q,q^{\prime},2}[m,\Tilde n,\Tilde m]$ and $H_{q,q^{\prime},3}[m,\Tilde n,\Tilde m]$ are respectively defined as follows.
	\begin{eqnarray}
		\label{Define_Hq_1}
		H_{q,q^{\prime},1}[m,\Tilde n,\Tilde m] \hspace{-3mm} &\Define& \hspace{-3mm} \sum_{i=1}^{p_{q}} h_{q,i} e^{-j2\pi \big(\Tilde m+\frac{M}{g_{1}}\lfloor q' / g_{2}\rfloor\big)\Delta f \tau_{q,i}} \nonumber \\ 
		 & & \hspace{-16mm}   e^{j2\pi \nu_{q,i} \Big(\Tilde n+\frac{N}{g_{2}}(q)_{g_{2}}\Big)T} \left[\frac{e^{j2\pi\xi \left(1-\frac{\tau_{q,i}}{T}\right)}-1}{j2\pi\xi }\right].
	\end{eqnarray}
	\vspace{-5mm}
	\begin{eqnarray}
		\label{Define_Hq_2}
		H_{q,q^{\prime},2}[m,\Tilde n,\Tilde m] \hspace{-3mm} &\Define& \hspace{-3mm} \sum_{i=1}^{p_{q}} h_{q,i}   e^{-j2\pi \big(\Tilde m+\frac{M}{g_{1}}\lfloor q' / g_{2}\rfloor\big)\Delta f \tau_{q,i}} e^{j2\pi\xi}  \nonumber \\ 
		& & \hspace{-14mm}  e^{j2\pi \nu_{q,i} \Big(\Tilde n-1+\frac{N}{g_{2}}(q)_{g_{2}}\Big)T} \left[\frac{1-e^{-j2\pi\xi\frac{\tau_{q,i}}{T}}}{j2\pi\xi}\right].
	\end{eqnarray} 
	\begin{eqnarray}
		\label{Define_Hq_3}
		H_{q,q^{\prime},3}[m,\Tilde n,\Tilde m] \hspace{-3mm} &\Define& \hspace{-3mm} \sum_{i=1}^{p_{q}} h_{q,i}  e^{-j2\pi \big(\Tilde m+\frac{M}{g_{1}}\lfloor q' / g_{2}\rfloor\big)\Delta f \tau_{q,i}} e^{j2\pi\xi} \nonumber \\ 
		& & \hspace{-23mm} e^{j2\pi \nu_{q,i} \Big(\frac{N}{g_{2}}\left((q)_{g_{2}}+1\right)-1\Big)T}  \left[\frac{1-e^{-j2\pi\xi\frac{\tau_{q,i}}{T}}}{j2\pi\xi}\right] \delta[\Tilde n].
	\end{eqnarray}
	
    \vspace{-2mm}
	\section{Proof of Theorem \ref{Proposed_DDRG_rect_lemma_2}}
	\label{proof_of_lemma_2}
		By substituting the expression of $Y_{q^{\prime}}[\Tilde n,\Tilde m]$ from (\ref{Rx_TFRG_simplified_exp_rect}) in (\ref{SFFT_proposed}), we get a resultant expression of $y_{q^{\prime}}[{\Tilde u}, {\Tilde v}]$ in terms of
		$X_{q}[ n,  m]$. Substituting the
		expression for $X_{q}[ n,  m]$ from (\ref{newInvSFFT}) in this resultant expression of $y_{q^{\prime}}[{\Tilde u}, {\Tilde v}]$ we get (\ref{Pro_DDRG_rect_expsion}) where $h_{q,q',1}[\Tilde{u},\Tilde{v},u,v], h_{q,q',2}[\Tilde{u},\Tilde{v},u,v]$ and $w_{q^{\prime}}[\Tilde u, \Tilde v]$ are respectively defined as follows.
	\vspace{-2mm}
	\begin{eqnarray}
    	\label{define_h_qq_1}
    	\nonumber h_{q,q',1}[\Tilde{u},\Tilde{v},u,v] \hspace{-3mm} &\Define& \hspace{-3mm} \frac{1}{MN}\hspace{-1mm}\sum_{i=1}^{p_q}h_{q,i}e^{j2\pi \left(\frac{N\nu_{q,i}}{g_{2}\Delta f}(q)_{g_{2}}-\frac{M\tau_{q,i}}{g_{1}T}\left\lfloor \frac{q^{\prime}}{g_{2}}\right\rfloor\right)} \nonumber \\
        & & \hspace{-27mm} \sum_{m=0}^{\frac{M}{g_{1}}-1}\hspace{-1mm} e^{-j2\pi m \left(\frac{(q)_{g_{1}}+g_{1}v}{M}\right)}\hspace{-1mm}\sum_{\Tilde{m}=0}^{\frac{M}{g_{1}}-1}  \hspace{-1mm} e^{j2\pi \Tilde{m}\left(\frac{g_{1}\Tilde{v}}{M}-\frac{\tau_{q,i}}{T}\right)}\Bigg\{\hspace{-2mm} \left(1-\frac{\tau_{q,i}}{T}\right)\nonumber \\ 
    	& & \hspace{-27mm}  e^{j\pi\xi\left(1-\frac{\tau_{q,i}}{T}\right)}\operatorname{sinc}\left(\xi\left(1-\frac{\tau_{q,i}}{T}\right)\right)\Bigg(\frac{N}{g_{2}}e^{j\pi\left(\frac{N}{g_{2}}-1\right)\psi} \nonumber \\
    	& & \hspace{-27mm} \left.\frac{\operatorname{sinc}(N\psi/g2)}{\operatorname{sinc}(\psi)}\right)\hspace{-1mm}+\hspace{-1mm} \left[\left(\frac{N}{g_{2}}e^{j\pi\left(\frac{N}{g_{2}}-1\right)\psi}\frac{\operatorname{sinc}(N\psi/g2)}{\operatorname{sinc}(\psi)}\right)-1\right]  \nonumber \\
    	& & \hspace{-27mm} \frac{\tau_{q,i}}{T} \operatorname{sinc}\left(\frac{\xi\tau_{q,i}}{T}\right) e^{-j2\pi \left(\frac{\lfloor q/g_{1}\rfloor+g_{2}u}{N}+\frac{\nu_{q,i}}{\Delta f}+\xi\left(\frac{\tau_{q,i}}{2T}-1\right)\right)}  \Bigg\}.
    \end{eqnarray}
    \vspace{-4mm}
    \begin{eqnarray}
    	\label{define_h_qq_2}
    	\nonumber h_{q,q',2}[\Tilde{u},\Tilde{v},u,v] \hspace{-3mm} &\Define& \hspace{-3mm}  \frac{ 1}{MN}\sum_{i=1}^{p_q}h_{q,i}e^{j2\pi \left(\frac{N}{g_{2}}-1\right)\left(\frac{\lfloor q/g_{1}\rfloor+g_{2}u}{N}+\frac{\nu_{q,i}}{\Delta f}\right)}  \nonumber \\
    	& & \hspace{-25mm} \frac{\tau_{q,i}}{T}e^{j2\pi \left(\frac{N\nu_{q,i}}{g_{2}\Delta f}(q)_{g_{2}}-\frac{M\tau_{q,i}}{g_{1}T}\lfloor q' / g_{2}\rfloor\right)} \sum_{\Tilde{m}=0}^{\frac{M}{g_{1}}-1}  e^{j2\pi \Tilde{m}\left(\frac{g_{1}\Tilde{v}}{M}-\frac{\tau_{q,i}}{T}\right)}  \nonumber \\
    	& & \hspace{-23mm} \sum_{m=0}^{\frac{M}{g_{1}}-1} 
    	 e^{-j2\pi m \left(\frac{(q)_{g_{1}}+g_{1}v}{M}\right)}  e^{j2\pi\xi\left(1-\frac{\tau_{q,i}}{2T}\right)} \operatorname{sinc}\left(\frac{\xi\tau_{q,i}}{T}\right).
    \end{eqnarray}
    \vspace{-3mm}
    \begin{eqnarray}
		\label{noise_rect_DDRG}
		w_{q^{\prime}}[\Tilde u, \Tilde v] \hspace{-3mm} &\Define& \hspace{-3mm}  	\sum_{\Tilde{n}=0}^{\frac{N}{g_{2}}-1}\sum_{\Tilde{m}=0}^{\frac{M}{g_{1}}-1}W_{q^{\prime}}[\Tilde n, \Tilde m]e^{j2\pi \left(\frac{\Tilde{m}\Tilde{v}}{(M/g_{1})}-\frac{\Tilde{n}\Tilde{u}}{(N/g_{2})}\right)}.
	\end{eqnarray}
	Here, $\xi$ is defined in (\ref{integral_rect}), and $\psi$ is given by
	\begin{eqnarray*}
		\label{define_Theta_1_2}
		\psi &\Define& \frac{\lfloor q/g_{1} \rfloor+g_{2}(u-\Tilde u)}{N}+\frac{\nu_{q,i}}{\Delta f}.
	\end{eqnarray*}

	\section{Proof of Lemma \ref{lemma3_Interlvd_TF}}
	\label{proof_of_lemma3_Doplr_1}
	Substituting the expression of $r(t)$ from (\ref{Rx_TD_signal_1}) in the R.H.S. of (\ref{Rx_TF_signal}) and using (\ref{rect_pulses}) we get
	\begin{eqnarray}
		\label{TFRG_rect_Doplr_2}
		Y[ \Tilde{n}, \Tilde{m}] \hspace{-1mm}&=&\hspace{-1mm} \frac{1}{\sqrt{T}}\sum_{q=0}^{Q-1}\sum_{i=1}^{p_q}\sum_{m=0}^{M-1}\sum_{n=0}^{N-1} h_{q,i}X_{q}[n,m]  \nonumber \\
		& & \hspace{-9mm} F_{q}[n,m, \Tilde{n}, \Tilde{m}]e^{-j2\pi \left(m \Delta f+\nu_{q,i}\right) \tau_{q,i}} + W[\Tilde n, \Tilde m], \nonumber \\
		F_{q}[n,m, \Tilde{n}, \Tilde{m}]\hspace{-2mm} &\Define& \hspace{-2mm} \int_{ \Tilde{n}T}^{\left( \Tilde{n}+1\right)T} \hspace{-9mm} g_{tx}\big(t-\tau_{q,i}-nT\big) e^{j2\pi \xi_{1} \frac{t}{T}} dt, \nonumber \\
		\mbox{\small{where}},\,\,\xi_{1} \hspace{-3mm} &\Define& \hspace{-3mm} \left(\left(m- \Tilde{m}\right) + T\nu_{q,i}\right),
	\end{eqnarray}
	\begin{eqnarray}
	\label{noise_GBMA}
			W[\Tilde n, \Tilde m] \hspace{-3mm} &\Define& \hspace{-3mm} \frac{1}{\sqrt{T}} \int_{ \Tilde{n}T}^{\left( \Tilde{n}+1\right)T} w(t) e^{-j2\pi  \Tilde{m}\Delta f t} dt.  
	\end{eqnarray}
	After simplifying the integral in the R.H.S. of the expression of $F_{q}[n,m, \Tilde{n}, \Tilde{m}]$ in (\ref{TFRG_rect_Doplr_2}) we get
	\begin{eqnarray}
		\label{Solve_Doplr_TFRG_1}
		F_{q}[n,m, \Tilde{n}, \Tilde{m}] \hspace{-3mm}&=&\hspace{-3mm} e^{j2\pi \xi_{1} \frac{\tau_{q,i}}{T}}e^{j2\pi n \nu_{q,i}T} \nonumber \\
		& &  \int_{t_{l}}^{t_u} g_{tx}\big(t\big) e^{j2\pi \left(\left(m-\Tilde{m}\right) \Delta f+\nu_{q,i}\right)t} dt 
	\end{eqnarray}
	where $t_{l} = \left(\Tilde{n}-n\right)T-\tau_{q,i}$ and $t_{u} =  \left(\Tilde{n}-n+1\right)T-\tau_{q,i}$. The integral in (\ref{Solve_Doplr_TFRG_1}) is non-zero only for $n=\Tilde n$ and $n= \Tilde n -1$. If $n=\Tilde n$, then it suffices
	to have the lower and upper limits for the integral as $0$ and $T-\tau_{q,i}$ respectively since $0 \leq \tau_{q,i} < T$. If $n=\Tilde n-1$, then it suffices
	to have the lower and upper limits for the integral as $T-\tau_{q,i}$ and $T$ respectively. On substituting these conditions in (\ref{Solve_Doplr_TFRG_1}), the expression of $F_{q}[n,m, \Tilde{n}, \Tilde{m}]$ is given by
	\begin{eqnarray}
		\label{integral_solution_doplr}
		F_{q}[n,m, \Tilde{n}, \Tilde{m}] \hspace{-3mm}&=&\hspace{-3mm}  e^{j2\pi \xi_{1} \frac{\tau_{q,i}}{T}}e^{j2\pi n \nu_{q,i}T} \nonumber \\
		& & \hspace{-12mm} \Bigg\{\sqrt{T}  \left[\frac{e^{j2\pi  \xi_{1}\left(1-\frac{\tau_{q,i}}{T}\right)}-1}{j2\pi  \xi_{1}}\right]\delta[n-\Tilde n] + \sqrt{T} e^{j2\pi  \xi_{1}} \nonumber \\
		& & \hspace{-5mm} \left[\frac{1-e^{-j2\pi  \xi_{1}\frac{\tau_{q,i}}{T}}}{j2\pi  \xi_{1}}\right] \delta[n-(\Tilde n -1)]\Bigg\}.
	\end{eqnarray}
	Substituting the expression of $F_{q}[n,m, \Tilde{n}, \Tilde{m}]$ from (\ref{integral_solution_doplr}) into the R.H.S. of the expression of $Y[\Tilde n, \Tilde m]$ in (\ref{TFRG_rect_Doplr_2}) we get (\ref{TF_interlvd_l3}), where $\ddot{H}_{q,1} [m,\Tilde n, \Tilde m]$ and $\ddot{H}_{q,2} [m,\Tilde n, \Tilde m]$ are defined as follows. 
	\begin{eqnarray}
	\label{H_q_1_doppler}
			\ddot{H}_{q,1} [m,\Tilde n, \Tilde m] \hspace{-3mm}&\Define&\hspace{-3mm} \sum_{i=1}^{p_q} h_{q,i} e^{-j2\pi \Tilde{m} \Delta f \tau_{q,i}} e^{j2\pi \Tilde{n} \nu_{q,i}T} \nonumber \\
			& &  \qquad \qquad \left[\frac{e^{j2\pi  \xi_{1}\left(1-\frac{\tau_{q,i}}{T}\right)}-1}{j2\pi  \xi_{1}}\right].
	\end{eqnarray}
	\vspace{-5mm}
	\begin{eqnarray}
		\label{H_q_2_doppler}
		\ddot{H}_{q,2} [m,\Tilde n, \Tilde m] \hspace{-3mm}&\Define&\hspace{-3mm} \sum_{i=1}^{p_q} h_{q,i} e^{-j2\pi \Tilde{m} \Delta f \tau_{q,i}} e^{j2\pi (\Tilde{n}-1) \nu_{q,i}T} \nonumber \\
		& & \qquad  e^{j2\pi  \xi_{1}} \left[\frac{1-e^{-j2\pi  \xi_{1}\frac{\tau_{q,i}}{T}}}{j2\pi  \xi_{1}}\right].
	\end{eqnarray}
	\section{Proof of Theorem \ref{lema_interlevd}}
	\label{proof_of_lemma_interlevd}
		By substituting the expression of $Y_{q^{\prime}}[\Tilde n,\Tilde m]$ from (\ref{TF_interlvd_l3}) in (\ref{SFFT_Inter}), we get a resultant expression of $y_{q^{\prime}}[{\Tilde k}, {\Tilde l}]$ in terms of
	$X_{q}[ n, m]$. Substituting the
	expression for $X_{q}[ n, m]$ from (\ref{Interlevd_TF}) in this resultant expression of $y_{q^{\prime}}[{\Tilde k}, {\Tilde l}]$ we get (\ref{lema_int_exp}) where $\ddot{h}_{q,q'}[\Tilde{k},\Tilde{l},k',l']$ is given by
	
	{\vspace{-2mm}
	\small
	\begin{eqnarray}  
	\label{h_mat_interlevd}
	\ddot{h}_{q,q'}[\Tilde{k},\Tilde{l},k^{\prime},l^{\prime}]  \hspace{-3mm} &\Define& \hspace{-3mm} \frac{1}{MN}e^{j2\pi \left(\frac{(q^{\prime})_{g_3}  \Tilde{l}}{M}-\frac{\Tilde{k}}{N}\left\lfloor \frac{q^{\prime}}{g_3} \right\rfloor\right)}  \sum_{i=1}^{p_q} h_{q,i}\nonumber \\
	& & \hspace{-25mm} e^{j2\pi\left(\left(\frac{\nu_{q,i}}{\Delta f}+\frac{k'}{N}\right)\left\lfloor \frac{q^{\prime}}{g_3} \right\rfloor-\frac{(q^{\prime})_{g_3} \tau_{q,i}}{T}\right)} \sum_{n^{\prime}=0}^{\frac{N}{g_4}-1}e^{j2\pi n^{\prime}g_4\left(\frac{\nu_{q,i}}{\Delta f}+\frac{k^{\prime}-\Tilde k}{N}\right)} \nonumber \\
	\nonumber & & \hspace{-25mm} \sum_{m=0}^{M-1} e^{-j2\pi\frac{ml'}{M}}e^{j\pi \frac{((q)_{g_3}-m)(g_3-1)}{g_{3}}}\frac{ g_{3} \sinc \left((q)_{g_3}-m\right) }{\sinc \left(\frac{(q)_{g_3}-m}{g_3}\right)}  \sum_{m^{\prime}=0}^{\frac{M}{g_3}-1}  \nonumber \\ 
	& & \hspace{-25mm} e^{-j2\pi m^{\prime}g_3 \left(\frac{\tau_{q,i}}{T}-\frac{\Tilde l}{M}\right)}  \Bigg\{ e^{j\pi \frac{\epsilon(g_4-1)}{g_{4}}}\frac{ g_{4} \sinc \left(\epsilon\right) }{\sinc \left(\frac{\epsilon}{g_4}\right)} \nonumber \\
	\nonumber & & \hspace{-20mm} \left[\frac{e^{j2\pi \kappa\left(1-\frac{\tau_{q,i}}{T}\right)}-1}{j2\pi \kappa}\right] +   e^{j\pi \frac{(\epsilon+1)(g_4-1)}{g_{4}}}\frac{ g_{4} \sinc \left(\epsilon+1\right) }{\sinc \left(\frac{\epsilon+1}{g_4}\right)} \nonumber \\
	& & \hspace{-15mm} e^{-j2\pi \left(\frac{\nu_{q,i}}{\Delta f}+\frac{k^{\prime}}{N}\right)} e^{j2\pi \kappa} \left[\frac{1-e^{-j2\pi \kappa\frac{\tau_{q,i}}{T}}}{j2\pi \kappa}\right]\Bigg\}
	\end{eqnarray}
	where $\kappa \Define \left(\left(m-((q^{\prime})_{g_3}  + m^{\prime}g_3)\right) + T\nu_{q,i}\right)$, $\epsilon \Define \left(\lfloor q / g_3\rfloor-(\lfloor q^{\prime}/g_3 \rfloor + n^{\prime}g_4)\right)$ 	
and $w_{q^{\prime}}[\Tilde k, \Tilde l]$ is given by
	\begin{eqnarray} 
	\label{noise_int}
	w_{q^{\prime}}[\Tilde{k},\Tilde{l}] \hspace{-3mm} &\Define& \hspace{-3mm}   e^{j2\pi \left(\frac{(q^{\prime})_{g_3}  \Tilde{l}}{M}-\frac{\lfloor q^{\prime}/g_3 \rfloor\Tilde{k}}{N}\right)} \sum_{n^{\prime}=0}^{\frac{N}{g_4}-1}\sum_{m^{\prime}=0}^{\frac{M}{g_3}-1} W_{noise}\left[\lfloor q^{\prime}/g_3 \rfloor \right. \nonumber \\
	& & \hspace{3mm}\left. + n^{\prime}g_4,(q^{\prime})_{g_3}  + m^{\prime}g_3\right]  e^{j2\pi \left(\frac{m^{\prime}\Tilde{l}}{M/g_3}-\frac{n^{\prime}\Tilde{k}}{N/g_4}\right)}.
	\end{eqnarray}
	\normalsize}
     \vspace{-4mm} 
	\section{Proof of Theorem \ref{Lemma4_Doppler_DD}}
	\label{proof_of_Doplr_DDRG_lemma} 
	The received TF symbols $Y[\Tilde n,\Tilde m]$ (i.e., (\ref{TF_interlvd_l3}) with $X_q[n,m]$ given by (\ref{newInvSFFT_Doppler}))
	are converted to the DD domain by using SFFT (see (\ref{Rx_DD_signal})). 
	The resultant expression of the received DD domain symbols (denoted here by $y[\Tilde{k},\Tilde{l}]$) depends on the transmitted TF domain symbols $X_{q}[ n, m]$.
	Substituting the expression of $X_{q}[ n,  m]$ from (\ref{newInvSFFT_Doppler}) into the resultant expression of $y[\Tilde{k},\Tilde{l}]$ gives
	\begin{eqnarray}
		\label{Doplr_DDRG_prf_2}
		\hspace{-3mm} y[\Tilde{k},\Tilde{l}] \hspace{-3mm} &=& \hspace{-3mm}  \sum_{q=0}^{Q-1}\sum_{l=0}^{M-1}\sum_{k=q\cdot \frac{N}{Q}}^{\left(\frac{N}{Q}-G-1\right)+q\cdot \frac{N}{Q}} \hspace{-8mm}\Tilde x_{q}[k,l] \Tilde h_{q,q'}[\Tilde{k},\Tilde{l},k,l]\hspace{-1mm} + \hspace{-1mm} w[\Tilde{k},\Tilde{l}]
	\end{eqnarray}
	where
 	\begin{eqnarray}
 	\label{define_h_q_doplr}
 	\nonumber \Tilde h_{q,q'}[\Tilde{k},\Tilde{l},k,l] \hspace{-3mm} &\Define& \hspace{-3mm} \frac{1}{MN}\sum_{i=1}^{p_q} h_{q,i}\sum_{\Tilde{m}=0}^{M-1}\sum_{m=0}^{M-1}    e^{j2\pi\Tilde{m} \left(\frac{\Tilde{l}}{M}- \frac{\tau_{q,i}}{T}\right)}    \\
 	\nonumber & & \hspace{-23mm} e^{-j2\pi \frac{ml}{M}}e^{j\pi\left(\frac{k-\Tilde{k}}{N}+\frac{\nu_{q,i}}{\Delta f}\right)\left(N-1\right)}\left[\frac{N\operatorname{sinc}\left({k-\Tilde{k}}+\frac{N\nu_{q,i}}{\Delta f}\right)}{\operatorname{sinc}\left(\frac{k-\Tilde{k}}{N}+\frac{\nu_{q,i}}{\Delta f}\right)}\right] \nonumber \\
 	& & \hspace{-9mm} \Bigg\{\left(1-\frac{\tau_{q,i}}{T}\right)e^{j\pi\xi_{1}\left(1-\frac{\tau_{q,i}}{T}\right)} \operatorname{sinc}\left(\xi_{1}\left(1-\frac{\tau_{q,i}}{T}\right)\right)  \nonumber \\
 	& & \hspace{-9mm} + \frac{\tau_{q,i}}{T}e^{-j2\pi\frac{k}{N}} e^{-j\pi\xi_{1}\left(\frac{\tau_{q,i}}{T}\right)} \operatorname{sinc}\left(\frac{\xi_{1}\tau_{q,i}}{T}\right)\Bigg\},
 	\end{eqnarray} 
 	\vspace{-3mm}
 	\begin{eqnarray}
 	\label{w_k_l_Doplr}
			w[\Tilde{k},\Tilde{l}] &\Define& \sum_{\Tilde{n}=0}^{N-1}\sum_{\Tilde{m}=0}^{M-1} W[\Tilde{n},\Tilde{m}]e^{j2\pi\left(\frac{\Tilde{m}\Tilde{l}}{M}-\frac{\Tilde{n}\Tilde{k}}{N}\right)},
 	\end{eqnarray}
	where, $\xi_{1}$ is defined in (\ref{TFRG_rect_Doplr_2}). The $q'$-th UT is allocated the DDREs in the set $\mathcal{R}_{q^{\prime}}^{Dplr}$ and therefore, the received DD domain symbols for the $q'$-th UT
	is nothing but the expression of $y[\Tilde{k},\Tilde{l}]$ in (\ref{Doplr_DDRG_prf_2}) with $(\Tilde{k},\Tilde{l}) \in \mathcal{R}_{q^{\prime}}^{Dplr}$.
	\section{Proof of Theorem \ref{Lemma5_Delay_DD}}
 	\label{proof_of_Dly_DDRG_lemma}  
	By substituting the expression of $Y[\Tilde n,\Tilde m]$ from (\ref{TF_interlvd_l3}) in (\ref{Rx_DD_signal}), we get a resultant expression for the received DD domain symbols $y[\Tilde k, \Tilde l]$  in terms of the transmitted TF domain symbols $X_{q}[ n, m]$. Using the expression of $X_{q}[ n, m]$ from (\ref{newInvSFFT_Dly}) and substituting it in the resultant expression of $y[\Tilde k, \Tilde l]$ gives 
 	\begin{eqnarray}
 		\label{Dly_DDRG_proof_1}
 		\nonumber  y[\Tilde{k},\Tilde{l}] \hspace{-3mm} &=& \hspace{-3mm}  \sum_{q=0}^{Q-1}\sum_{k=0}^{N-1}\sum_{l=q\cdot \frac{M}{Q}}^{\left(\frac{M}{Q}-G-1\right)+q\cdot \frac{M}{Q}}\hspace{-7mm} \Tilde x_{q}[k,l] \hat{h}_{q,q'}[\Tilde{k},\Tilde{l},k,l] + w_{q'}[\Tilde{k},\Tilde{l}]
	 	\end{eqnarray}
	 	\vspace{-4mm}
 		\begin{eqnarray}
 			\label{Dly_h_q_q_rect}
 		\nonumber \hat{h}_{q,q'}[\Tilde{k},\Tilde{l},k,l] \hspace{-3mm} &\Define& \hspace{-3mm} \frac{1}{MN}\sum_{i=1}^{p_q} h_{q,i}\sum_{\Tilde{m}=0}^{M-1}\sum_{m=0}^{M-1}    e^{j2\pi\Tilde{m} \left(\frac{\Tilde{l}}{M}- \frac{\tau_{q,i}}{T}\right)}    \\
 		\nonumber & & \hspace{-21mm} e^{-j2\pi \frac{ml}{M}}e^{j\pi\left(\frac{k-\Tilde{k}}{N}+\frac{\nu_{q,i}}{\Delta f}\right)\left(N-1\right)}\left[\frac{N\operatorname{sinc}\left({k-\Tilde{k}}+\frac{N\nu_{q,i}}{\Delta f}\right)}{\operatorname{sinc}\left(\frac{k-\Tilde{k}}{N}+\frac{\nu_{q,i}}{\Delta f}\right)}\right] \nonumber \\
 		& & \hspace{-9mm} \Bigg\{\left(1-\frac{\tau_{q,i}}{T}\right)e^{j\pi\xi_{1}\left(1-\frac{\tau_{q,i}}{T}\right)} \operatorname{sinc}\left(\xi_{1}\left(1-\frac{\tau_{q,i}}{T}\right)\right)  \nonumber \\
 		& & \hspace{-9mm} + \frac{\tau_{q,i}}{T}e^{-j2\pi\frac{k}{N}} e^{-j\pi\xi_{1}\left(\frac{\tau_{q,i}}{T}\right)} \operatorname{sinc}\left(\frac{\xi_{1}\tau_{q,i}}{T}\right)\Bigg\},
 		\end{eqnarray}
 		\vspace{-3mm}
 		\begin{eqnarray}
 		    \label{w_k_l_Dly}
 		    w[\Tilde{k},\Tilde{l}] &\Define& \sum_{\Tilde{n}=0}^{N-1}\sum_{\Tilde{m}=0}^{M-1} W[\Tilde{n},\Tilde{m}]e^{j2\pi\left(\frac{\Tilde{m}\Tilde{l}}{M}-\frac{\Tilde{n}\Tilde{k}}{N}\right)}.
 		\end{eqnarray}
	From the expression of $y[\Tilde{k},\Tilde{l}]$ in (\ref{Dly_DDRG_proof_1}), the received DD domain symbols of the $q^{\prime}$-th UT (i.e., $y_{q^{\prime}}[\Tilde{k},\Tilde{l}]$) at the BS is given by (\ref{Dly_DDRG_rect_expsion}). \vspace{-5mm}
	\end{appendices}

	\ifCLASSOPTIONcaptionsoff
	\newpage
	\fi


	
	%
	%

\end{document}